\documentclass[aps, reprint, twocolumn, superscriptaddress, citeautoscript]{revtex4-1}

\usepackage{graphicx}

\usepackage{amssymb}
\usepackage{amsmath}
\usepackage{amsfonts}
\usepackage{mathrsfs}

\usepackage{bm}
\usepackage{dcolumn}
\usepackage{color}
\usepackage[colorlinks=true,citecolor=blue]{hyperref}
\hypersetup{colorlinks=true,citecolor=blue,linkcolor=red, urlcolor=blue}

\usepackage{float}

\usepackage{color}
\usepackage{ulem}
\definecolor{purple}{rgb}{0.8,0,0.6}
\definecolor{orange}{rgb}{1,0.55,0}

\newcommand{\beqn}{\begin{eqnarray}}
	\newcommand{\eeqn}{\end{eqnarray}}

\newcommand{\comment}[1]{}

\begin{document}
	
	\title[]{Enhancement of Hot Carrier Effects and Signatures of Confinement in Terms of Thermalization Power in Quantum Well Solar Cells}
	
	\author{I. Makhfudz} 
	\affiliation{Aix Marseille Universit\'{e}, CNRS, Université de Toulon, IM2NP UMR 7334, 13397, Marseille, France}
	\author{N. Cavassilas}
	\affiliation{Aix Marseille Universit\'{e}, CNRS, Université de Toulon, IM2NP UMR 7334, 13397, Marseille, France}
	\author{M. Giteau}
	\affiliation{Research Center for Advanced Science and Technology, The University of Tokyo, Komaba 4-6-1, Meguro-ku,
		Tokyo 153-8904, Japan}\affiliation{NextPV, LIA RCAST-CNRS, The University of Tokyo, Komaba 4-6-1, Meguro-ku, Tokyo 153-8904, Japan}
	\author{H. Esmaielpour}
	\affiliation{NextPV, LIA RCAST-CNRS, The University of Tokyo, Komaba 4-6-1, Meguro-ku, Tokyo 153-8904, Japan}\affiliation{CNRS, Ecole Polytechnique-IP Paris, Institut Photovoltaique d'Ile de France (IPVF), UMR 9006, 18 Boulevard Thomas Gobert,
		91120 Palaiseau, France}
	\author{D. Suchet}
	\affiliation{NextPV, LIA RCAST-CNRS, The University of Tokyo, Komaba 4-6-1, Meguro-ku, Tokyo 153-8904, Japan}\affiliation{CNRS, Ecole Polytechnique-IP Paris, Institut Photovoltaique d'Ile de France (IPVF), UMR 9006, 18 Boulevard Thomas Gobert,
		91120 Palaiseau, France}
	\author{A.-M. Dar\'{e}}
	\affiliation{Aix Marseille Universit\'{e}, CNRS, Université de Toulon, IM2NP UMR 7334, 13397, Marseille, France}
	\author{F. Michelini}
	\affiliation{Aix Marseille Universit\'{e}, CNRS, Université de Toulon, IM2NP UMR 7334, 13397, Marseille, France}

	\address{}
	\vspace{10pt}
	
	\begin{abstract}
		A theoretical model using electron-phonon scattering rate equations is developed for assessing
		carrier thermalization under steady-state conditions in two-dimensional systems. The model
		is applied to investigate the hot carrier effect in III-V hot-carrier solar cells with a quantum well absorber. 
		The question underlying the proposed investigation is: what is the power required to
		maintain two populations of electron and hole carriers in a quasi-equilibrium state at fixed temperatures
		and quasi-Fermi level splitting? 
		The obtained answer is that the thermalization power density is reduced in two-dimensional systems compared to their bulk counterpart, which demonstrates a confinement-induced enhancement of the hot carrier effect in quantum wells. This power overall increases with the well thickness, and it is moreover shown that the intra-subband contribution dominates at small thicknesses while the inter-subband contribution increases with thickness and dominates in the bulk limit. 
		Finally, the effects of the thermodynamic state of phonons and screening are clarified. In particular, the two-dimensional thermalization power density exhibits a non-monotonic dependence on the thickness of the quantum well layer, when both out-of-equilibrium longitudinal optical phonons and screening effects are taken into account. 
		Our theoretical and numerical results provide tracks to interpret intriguing experimental observations in quantum well physics. They will also offer guidelines to increase the yield of photovoltaic effect
		based on the hot carrier effect using quantum well heterostructures, a result critical to the research toward high-efficiency solar cell devices.
	\end{abstract}
	
	%
	%
	%
	%
	%
	
	\maketitle
	


\section{Introduction\label{Intro.}}

In the contemporary solar cell research progress, it has been suggested that if the charge carriers of the semiconductor constituting the solar cell can be maintained at high effective temperature, or eventually in an "hot" a-thermal distribution, the yield of the solar cell can be improved~\cite{Conibeer1}. Major efforts and proposals have thus been put forward in order to achieve hot carrier effect~\cite{Conibeer2}. Hot carrier phenomenon has actually been known in semiconductor physics research since early in the 60's~\cite{JShah}. Its relevance to solar cell study was shown by Ross and Nozik~\cite{RossNozik} who first demonstrated the hot carrier effect as a principle to beat the efficiency limit imposed by Shockley-Queisser theory~\cite{ShockleyQueisser}.  

A key mechanism for achieving hot carrier effect is impeding the carrier thermalization process, that is, preventing the charge carriers from cooling down via their interaction with phonons~\cite{Mahan}. There are various ways in which this can happen which include reducing carrier-phonon coupling strength~\cite{JShahSolidStateElectronics}, reducing longitudinal optical (LO) phonon density~\cite{MahyarAPL}, and reducing LO phonon decay to acoustic phonons~\cite{RidleyBook}. 
Overall, electrons and holes under continuous illumination can reach a temperature higher than the one of lattice if the rate at which the power is dissipated via carrier-phonon scattering can be reduced. 

One of the most promising tracks to promote a hot carrier effect is to confine electrons in a small region, which restricts their interaction with phonons and, hence, prevents electrons from thermalizing. This can be achieved by employing semiconductor heterostructures with reduced dimensionality (D), in which carriers experience confinement in one (2D), two (1D) or three (0D) directions~\cite{Hirst0}, compared to the bulk case (3D). In such low dimensional systems, the band structure splits into subbands whose energy separation depends on semiconductor properties, as energy gap and effective mass in a simple empirical modeling.
In optoelectronics and photovoltaics, quantum wells (2D) based on  polar semiconductors of III-V family are widely used due to their good absorption properties. In these materials,  the predominant electron thermalization channel is the scattering between electrons and longitudinal optical (LO) phonons~~\cite{JShahSolidStateElectronics}, while optical phonons may themselves decay into acoustic phonons~\cite{ShahLeite}.
While several experiments have indicated that the approach does give rise to a hot carrier effect \cite{Pelouch,Ryan,ShahPRL,Lyon,Balkan,Rosenwaks,Hirst,ConibeerIEEE}, 
only a few theoretical investigations estimates the contribution of quantum confinement to the hot carrier effect.
Furthermore, understanding the intricate mechanisms of this contribution is key to improve the design of solar cells integrating nanostructured absorbers~\cite{ConibeerIEEE}.

Numerous theoretical studies on charge transport in quantum wells exist from works in the 80's, but mostly focused on the calculation of scattering rates without specific application to hot carrier phenomenon 
nor relevance to solar cells \cite{Ferry1978,Hess,Price1981,Ridley1,Ridley2}. 
These works reported that a quantum well structure increases the electron-phonon scattering rate relatively to its value in the bulk material.
Other works applying Boltzmann transport equation on ultrafast dynamics of electrons and phonons in a quantum well~\cite{LugliGoodnickPRL} however reported that, when the LO phonons are assumed to be in a non-thermal distribution,  the electron-phonon scattering rate is of the same order~\cite{GoodnickLugliPRB} or even reduced compared to the 3D case~\cite{Ge,Sarma}. 
The apparent opposite behaviors in terms of electron-phonon scattering rate in quantum wells with respect to bulk points out the critical role of LO phonons throughout the type of equilibrium or non-equilibrium distributions they follow. 
Finally, related theoretical studies demonstrated high effective electron and hole temperatures on short time scales upon ultrafast excitation and estimated the relaxation time from the transient decay of the effective temperature~\cite{JoshiFerry,GoodnickReview}.

In the present work, we demonstrate that confinement in a quantum well suppresses the electron-phonon scattering rate, a phenomenon referred to as the phonon bottleneck~\cite{Bastard}. This phenomenon has been observed in quantum dots where the discretization of energy levels prevents state relaxation~\cite{QDphononbottleneck}. As a consequence, the hot  carrier effect is shown to be enhanced from quantum confinement. For that, we discuss a quantity which can be deduced from experiment, the so-called thermalization power. 
Our work relies on a semi-classical kinetic theory developed recently for bulk solar cells~\cite{Tsai}, which we have extended to include 2D confinement for assessing carrier thermalization in quantum well solar cells.
Moreover, simulations are performed for InGaAsP/InGaAs/InGaAsP quantum wells varying the well thickness, and using input parameters from experiment performed under steady-state, \textit{i.e.}, non-transient excitation, thus close to realistic situations in which solar cells normally operate. 

This article begins with a description of the main lines of the phenomenological model of carrier cooling in homogeneous bulk solar cells. Then, a specific model of electron-phonon interaction is built taking into account the confinement of electrons  in a finite quantum well, and then used to derive the electron-phonon scattering rate, detailing the distinction between intra and inter-subband processes. 
The unconfined LO phonons in the model are characterized by a non-thermal distribution.
The role of a non-equilibrium state of LO phonons on carrier cooling is clarified from the scattering rate. The thermalization power needed to achieve electron and hole populations at given temperatures is computed and analyzed, with an emphasis on the relative importance of a non-equilibrium phonon distribution and screening effect in the case of confining heterostructures. Finally, our theoretical findings are discussed with respect to a number of experimental observations reported in the literature, old and recent, in the context of seeking for the hot carrier effect in quantum wells.   

\section{On Steady-State Carrier Thermalization and Thermalization Power}
\subsection{Principle}

Absorption of a photon generates a pair of electron and hole in the conduction and valence bands respectively of a semiconductor. For photon energy larger than the band gap, the electron (hole) may be generated with energy well above (below) the conduction (valence) band edge. The ensemble of photo-generated carriers redistribute the absorbed energy among themselves, and may form an effective high-temperature thermal distribution, referred as hot carrier effect. But, this excess kinetic energy is mainly dissipated through carrier-phonon scattering, in a process known as thermalization, until the carriers reach the band edges and fulfill an effective equilibrium distribution at the lattice temperature. The tendency of electrons to cool down can however be impeded if the electron-LO phonon scattering rate is strongly suppressed, so that electrons can hardly release their energy by emitting LO phonons. In other words, the net phonon emission rate needs to be reduced, either by reducing phonon emission or by increasing phonon absorption.  

The rate of energy transfer from electrons to LO phonons thus measures the tendency of electrons to cool down. The corresponding power is called thermalization power. It is the quantity of central interest in our work because of two important reasons; it can be deduced from  experiment and it is meaningful: a higher thermalization power for a given carrier temperature means that it is necessary to pump more energy into the system to achieve the same carrier temperature. 

\begin{figure}
	\includegraphics[width=\columnwidth]{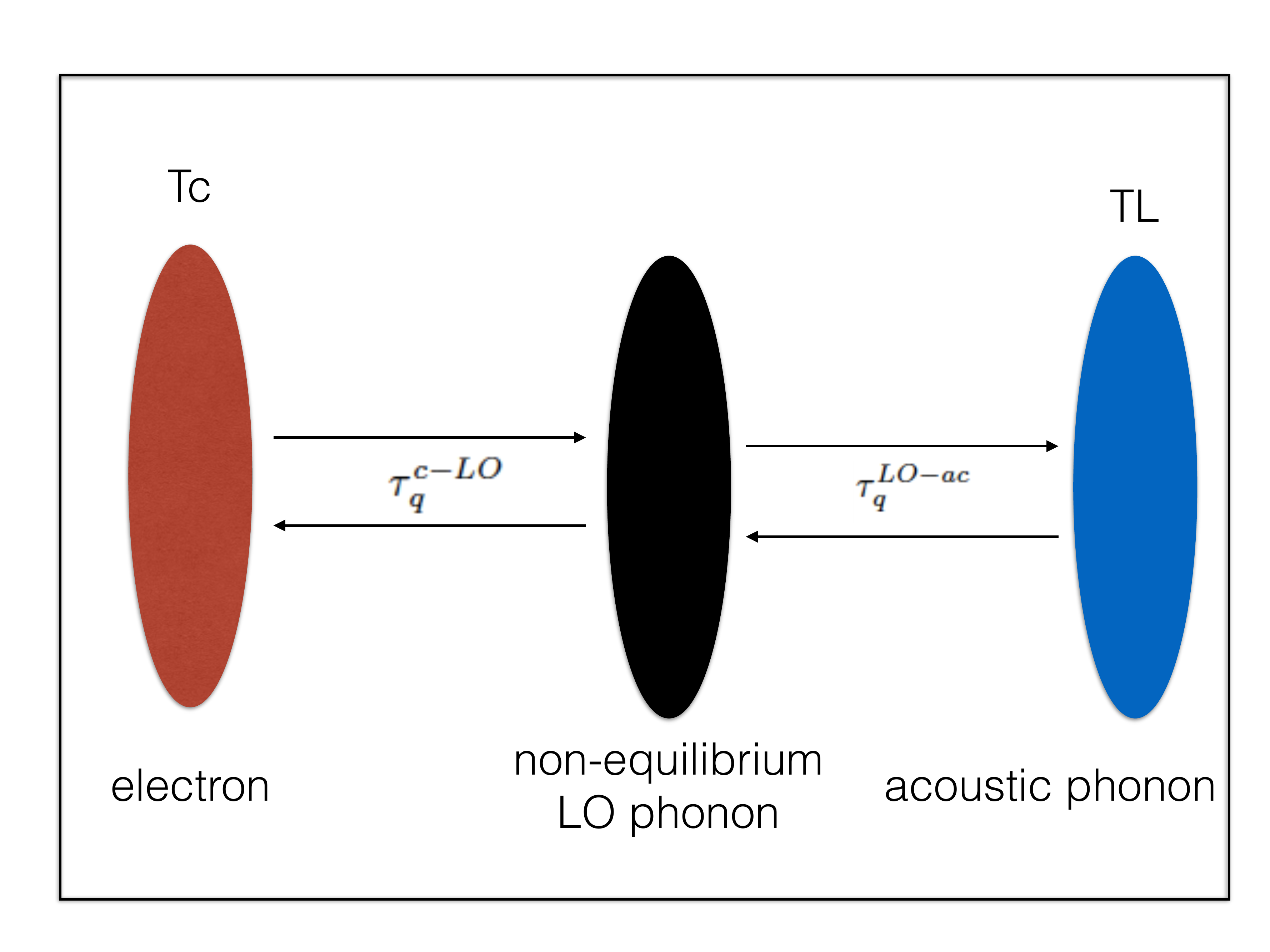}
	\label{fig:tsaimodel}
	\caption{(Color online) Illustration of a steady-state hot carrier effect involving non-equilibrium hot LO phonons.}
\end{figure}

\subsection{Main lines of 3D theoretical model}

A semi-classical model of carrier cooling was derived by Tsai in Ref.~\cite{Tsai} for homogeneous 3D systems of charge carriers, and longitudinal and acoustic phonons in bulk semiconductors used in solar cells. 
The fundamental assumption of the model is that acoustic phonons remain at lattice temperature $T_L$ while electrons form a thermalized population at a carrier temperature $T_c$ that can be significantly higher than $T_L$, thus becoming hot carriers. The first part of the assumption is justified by the fact that acoustic phonons do not interact directly with electrons because in polar semiconductors, electrons only couple strongly to polar optical phonons~\cite{JShahSolidStateElectronics,ShahLeite}. 
The thermalization power can be computed from the rate of change of the number of LO phonons due to their interaction with electrons, which represents the only path of thermalization for carrier in that model.
More precisely, the thermalization power in units of W/m$^2$ is given by
\begin{equation}\label{thermalpower3d}
	P^{c-LO}_{\mathrm{th}} = L^{ph}_z \int \frac{d^3\mathbf{q}}{(2\pi)^3} \hbar\omega_{\mathbf{q}} \frac{dN_{\mathbf{q}}}{dt}\Bigr|_{c-LO},
\end{equation}
where  $L^{ph}_z$ is the dimension of the LO phonon bath along the well confinement direction $\mathbf{q}$ is the LO phonon wave vector, $\hbar\omega_{\mathbf{q}}$ its energy, while $N_{\mathbf{q}}$ is the number of phonons at wave vector $\mathbf{q}$ which is not necessarily equal to its equilibrium value.  The equilibrium population at a given temperature $T$ is given by the Bose-Einstein distribution: $N_{\mathbf{q}}(T)=(\exp(\hbar\omega_{\mathbf{q}}/(k_BT))-1)^{-1}$ where $\hbar$ is the reduced Planck constant, and $k_B$ the Boltzmann constant.

A key insight put forward in Ref.~\cite{Tsai} argues that the total rate $dN_{\mathbf{q}}/dt$ for LO phonons receives two principal contributions: one from electrons which emit or absorb LO phonons and the other from the acoustic phonons from 3-phonon scattering process where an LO phonon transforms into two acoustic phonons or vice versa~\cite{RidleyBook}. In other words, we can write
\begin{equation}
	\frac{dN_{\mathbf{q}}}{dt}=\frac{dN_{\mathbf{q}}}{dt}\Bigr|_{c-LO}+\frac{dN_{\mathbf{q}}}{dt}\Bigr|_{LO-ac}  ,
\end{equation}
where
\begin{equation}\label{scatteringrateequationcLO}
	\frac{dN_{\mathbf{q}}}{dt}\Bigr|_{c-LO}=\frac{N_{\mathbf{q}}(T_c)-N_{\mathbf{q}}}{\tau^{c-LO}_{\mathbf{q}}} \ , 
\end{equation}
\begin{equation}\label{scatteringrateequationLOac}
	\frac{dN_{\mathbf{q}}}{dt}\Bigr|_{LO-ac}=\frac{N_{\mathbf{q}}(T_L)-N_{\mathbf{q}}}{\tau^{LO-ac}_{\mathbf{q}}} \ ,
\end{equation}
and the subscripts $c$, $LO$, $ac$ stand for charge carrier, longitudinal optical phonon, and acoustic phonon respectively, to be used in the rest of this article.

A steady-state is achieved when the rate of LO phonon emission by electrons in Eq.~(\ref{scatteringrateequationcLO}) equals the rate of the LO phonon conversion into acoustic phonons in Eq.~(\ref{scatteringrateequationLOac}). Under this condition, the LO phonons themselves will remain out of equilibrium, with a population number that can be expressed as a thermodynamic average between 
Bose-Einstein distributions at $T_c$ and $T_L$,  with no well-defined effective temperature. 
Setting $dN_{\mathbf{q}}/dt=0$, one arrives at~\cite{Tsai}
\begin{equation}\label{LOphononpopulationnumber}
	N_{\mathbf{q}}=\frac{N_{\mathbf{q}}(T_c)\tau^{LO-ac}_{\mathbf{q}}+N_{\mathbf{q}}(T_L)\tau^{c-LO}_{\mathbf{q}}}{\tau^{c-LO}_{\mathbf{q}}+\tau^{LO-ac}_{\mathbf{q}}} \ ,
\end{equation}
which defines an athermal population of LO phonons. Substituting Eq.~(\ref{LOphononpopulationnumber}) into Eqs.~(\ref{thermalpower3d}) and (\ref{scatteringrateequationcLO}), one obtains~\cite{Tsai} 
\begin{equation}\label{thermalpower3dbulk}
	P^{c-LO}_{\mathrm{th}}=L^{ph}_z\int \frac{d^3q}{(2\pi)^3} \hbar\omega_{\mathbf{q}} \frac{N_{\mathbf{q}}(T_c)-N_{\mathbf{q}}(T_L)}{\tau^{c-LO}_{\mathbf{q}}+\tau^{LO-ac}_{\mathbf{q}}} 
\end{equation}
for the thermalization power under steady-state condition. 
The thermalization power reflects the deexcitation of hot carriers via interaction with LO phonons. Thus a reduced thermalization power represents a lower cost to  
maintain a carrier population hot, it is hence beneficial for technological purposes. 

For 3D bulk system, the electron-LO phonon scattering rate is given by~\cite{Tsai}
\begin{equation}\label{bulkelectronphononscatteringrate}
	\frac{1}{\tau^{c-LO}_{\mathbf{q}}}=\frac{m^2_ck_BT_c|M_{\mathbf{q}}|^2V_c}{\pi\hbar^5q}\mathrm{ln}\left[\frac{1+\exp(\eta_c-\varepsilon_{\mathrm{min}}+\varepsilon_{\mathrm{LO}})}{1+\exp(\eta_c-\varepsilon_{\mathrm{min}})}\right] \ ,
\end{equation}
where $m_c$ is the carrier mass, $M_q$ the electron-phonon coupling function, $V_c$ the carrier spatial volume, 
$q$ the LO phonon wave vector. Furthermore $\eta_c=\mu_c/(k_BT_c)$, where $\mu_c$ is the chemical potential, $\varepsilon_{\mathrm{LO}}=\hbar\omega_{\mathrm{LO}}/(k_BT_c)$, $\omega_{\mathrm{LO}}$ the LO phonon frequency (assumed constant independent of $q$), $\varepsilon_{\mathrm{min}}=\hbar^2k^2_{\mathrm{min}}/(2m_ck_BT_c)$ where $k_{\mathrm{min}}=q/2+m_c\omega_{\mathrm{LO}}/(\hbar q)$.

To show that non-equilibrium phonons do contribute to hot carrier effect, we compare Eq.~(\ref{thermalpower3d}) with the corresponding expression assuming LO phonons to be at equilibrium at lattice temperature $T_L$. Then $N_{\mathbf{q}}=N_{\mathbf{q}}(T_L)$. Substituting it into Eq.~(\ref{scatteringrateequationcLO}) gives
\begin{equation}\label{thermalpower3dequilibriumLOphonons}
	P^{c-LO}_{\mathrm{th}}=L^{ph}_z\int \frac{d^3\mathbf{q}}{(2\pi)^3} \hbar\omega_{\mathbf{q}} \frac{N_{\mathbf{q}}(T_c)-N_{\mathbf{q}}(T_L)}{\tau^{c-LO}_{\mathbf{q}}} \ ,
\end{equation}
which is always larger than the thermalization power with non-equilibrium phonons expressed as in Eq.~(\ref{thermalpower3dbulk}). This thus demonstrates explicitly that non-equilibrium hot phonons help to maintain a hot carrier effect in bulk systems. It is emphasized here that the hot carrier effect studied in our work is a steady-state property, thus distinguishing itself from the transient hot carrier effect known in the literature~\cite{GoodnickReview}.

The analytical expression for the electron-LO phonon scattering time $\tau^{c-LO}_{\mathbf{q}}$  (Eq.~(\ref{bulkelectronphononscatteringrate})) was derived in Ref.~\cite{Tsai} for bulk homogeneous semiconductor solar cells. In the following section, we will derive the expression for electron-LO phonon scattering time for 2D systems, and we will analyze the effect of confinement on the hot carrier effect.


\section{2D Theoretical Model of Confined Carrier Thermalization}
In this part, we propose to derive expressions for the electron-LO phonon scattering rate $\tau^{c-LO}_{\mathbf{q}_{\perp},m}$ and the resulting thermalization power $P^{c-LO}_{\mathrm{th}}$ for two-dimensional semiconductor heterostructures.
The general framework of the derivation assumes that electron-electron scattering is faster than any other process, such that the carriers follow a Fermi-Dirac distribution given by a temperature $T_c$ hotter than the lattice one, and a quasi Fermi level $\mu_c$.
The expression for 2D systems starts from Eq.~(\ref{thermalpower3d})
\begin{eqnarray}\label{thermalpower2d}
	P^{c-LO}_{\mathrm{th}} &=&\sum_m\int \frac{d^2\mathbf{q}_{\perp}}{(2\pi)^2} \hbar\omega_{\mathbf{q}_{\perp},m} \frac{dN_{\mathbf{q}_{\perp},m}}{dt}|_{c-LO}\nonumber \\
	&=&\sum_m\int \frac{d^2\mathbf{q}_{\perp}}{(2\pi)^2} \hbar\omega_{\mathbf{q}_{\perp},m} \nonumber \\
	&&    \times \frac{N_{\mathbf{q}_{\perp},m}(T_c)-N_{\mathbf{q}_{\perp},m}(T_L)}{\tau^{c-LO}_{\mathbf{q}_{\perp},m}+\tau^{LO-ac}_{\mathbf{q}_{\perp},m}} \ ,
\end{eqnarray}
which involves a sum over the number of phonon modes $m$, and an integration over the transverse wave vector $\mathbf{q}_{\perp}$ of the LO phonon. 
In the following treatment, the LO phonons are supposed dispersionless, such that we shall take $\hbar \omega_{\mathbf{q}_{\perp},m} = \hbar \omega_{LO}$, hence the only quantity in the integral which depends on $({\bf q}_\perp, m)$ is $\tau^{c-LO}_{{\bf q}_\perp, m}$.

\begin{figure}
	\includegraphics[width=\columnwidth]{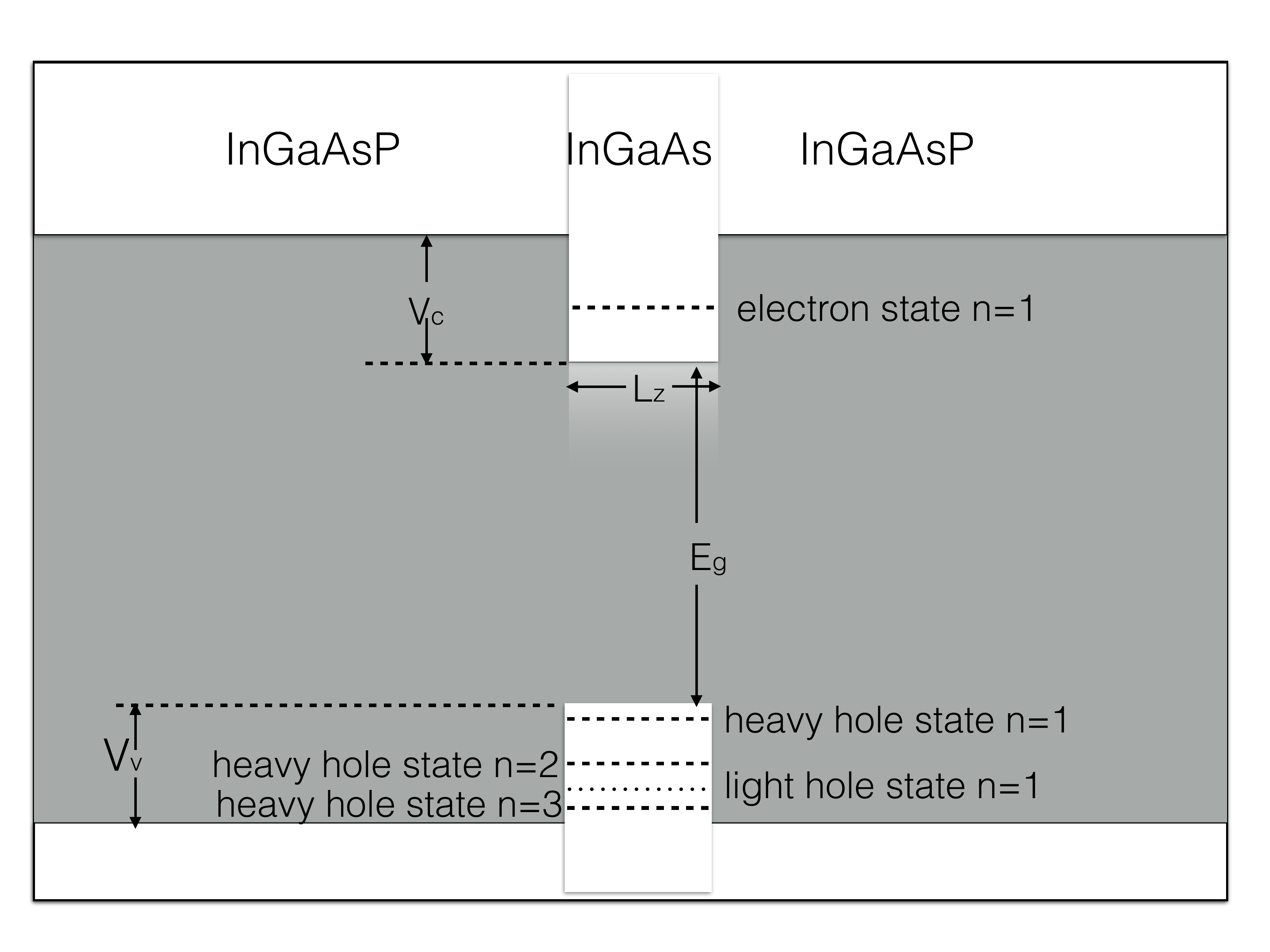}
	\caption{\label{fig:fig2} Illustration of the band diagram of a finite quantum well potential of InGaAs layer hosting a few confined states of electron and holes. For the calculations, $E_g=0.8$ eV,$V_{\mathrm{con}}=0.235$ eV, and $V_{\mathrm{val}}=0.185$ eV for the energy gap, the conduction band well depth, and valence band well depth respectively. All other parameters used in our calculations are given in Supp. Mat.}.
\end{figure}

To be concrete and perform illustrative calculations, we focus on a single quantum well of width $L_z$, made of an InGaAs layer sandwiched between two InGaAsP layers, as illustrated in Fig.~\ref{fig:fig2}.
In the analytical derivation, we will rely on the approximation of infinite barrier height for the wave functions that will be used to determine interaction matrix elements.
In the numerical calculations, we will consider finite barrier height for the confinement energies implemented to account for subband positions and fillings. This impacts the number of subbands existing in the well, as well as their minima (subband edge energy).

For the scattering between LO and acoustic phonons, we will use the following result~\cite{RidleyBook}
\begin{equation}\label{LOPacousticscatteringrate}
	\frac{1}{\tau^{LO-ac}_{\mathbf{q}}}=\frac{\Gamma^2\hbar\omega^3_{\mathbf{q}}(N_{\mathbf{q}_1}+N_{\mathbf{q}_2}+1)}{32\pi\rho v^3_s}=\frac{N_{\mathbf{q}_1}+N_{\mathbf{q}_2}+1}{\tau^{LO-ac}_0} \ .
\end{equation}
In this expression, $N_{\mathbf{q}_{1(2)}}$ is the population of acoustic phonons at $T_L$ at wave vector $\mathbf{q}_{1(2)}$ satisfying $\mathbf{q}=\mathbf{q}_1+\mathbf{q}_2$.
$\rho$ is the mass density of the material, $v_s$ the speed of sound wave, $\Gamma$ is a deformation potential coefficient related to the material's Gruneisen parameter by~\cite{RidleyBook}
\begin{equation}\label{eq:gamma}
	\Gamma=\sqrt{\frac{4}{3}}\frac{\gamma\omega_{q_{\perp},m}}{v_s} \ ,
\end{equation}
giving the zero temperature LO phonon-acoustic phonon scattering time
\begin{equation}\label{eq:tauLO-ac_0}
	\tau^{LO-ac}_0=\frac{32\pi\rho v^3_s}{\Gamma^2\hbar\omega^3_{q_{\perp},m}}.
\end{equation}

In the 3-phonon scattering process that converts an LO phonon into two acoustic phonons, the energy conservation law mandates that $\hbar\omega_{\mathbf{q}}=\hbar\omega_{\mathbf{q}_1}+\hbar\omega_{\mathbf{q}_2}$. While there are virtually infinitely many possible pairs of ($\mathbf{q}_1,\mathbf{q}_2$) that satisfy the above energy conservation equation for each given $\omega_{\mathbf{q}}$, we consider a simplifying assumption of equal energy partition normally employed in literature: $\hbar\omega_{\mathbf{q}_1}=\hbar\omega_{\mathbf{q}_2}=\hbar\omega_{\mathbf{q}}/2$~\cite{Klemens}.
It  gives the following population numbers of acoustic phonons 
\begin{equation}\label{BEacoustic}
	N_{\mathbf{q}_1}=N_{\mathbf{q}_2}=\frac{1}{\exp(\frac{\hbar\omega_{\mathbf{q}}}{2k_BT_L})-1}.
\end{equation}
Acoustic phonons are at equilibrium at $T_L$, in addition, they are not confined.  Thus, Eqs.~(\ref{LOPacousticscatteringrate}-\ref{BEacoustic}) will be used in the forthcoming 2D and 3D numerical calculations with 
\begin{equation}
	\tau^{LO-ac}_{\bf q} =  \tau^{LO-ac}_0 \frac{e^{\frac{\hbar \omega_{LO}}{2 k_B T_L}}-1}{e^{\frac{\hbar \omega_{LO}}{2 k_B T_L}}+1} .
	\label{eq:tauLO-ac}
\end{equation}
For InGaAs quantum well, $\Gamma$ of Eq.~(\ref{eq:gamma}) can be estimated from experimental values for InAs and GaAs~\cite{Sparks}.  For $T_L=300$~K, we thus have $\tau^{LO-ac}_0 =2.2 \ 10^{-11}$~s from Eq.~(\ref{eq:tauLO-ac_0}), and $\tau^{LO-ac}_{\bf q} \simeq 0.74 \ 10^{-11}$~s from Eq.~(\ref{eq:tauLO-ac}) (see Table I of Supp. Mat. for parameter values).

\subsection{Confined electron and unconfined phonon states}

In the proposed derivation, we will differ from the models developed for quantum well heterostructures in the 80's~\cite{Rucker91,Babiker,RidleyPRB,RidleyJPC,Huang} which considered the limit where both electrons and phonons are perfectly confined in the quantum well, assumed to be an infinite potential well in first approximation.
Instead, we propose to consider unconfined phonon states with several arguments. First of all, we rely on  the conclusion put foward in Ref.~\cite{Rucker91} that the impact of dimensionality in electron-LO phonon scattering rate so dramatically depends on the model used to account for phonon confinement that it is safer to use bulk phonon states. 
Moreover, such a simplified treatment has been widely adopted in theoretical literature~\cite{Price1981,dasSarmaMason}.
Experimentally,  it is reasonably realized in lattice-matched systems, where phonons are generally assumed to keep their bulk properties.
Finally, recent theoretical works~\cite{ZhangDysonRidley} which compared model based on bulk-like phonon and more sophisticated approach based on dielectric continuum model~\cite{RidleyBook2} where phonons can be confined also show that the difference in the result is relatively insignificant. 

The corner stone of the thermalization power defined in Eq.~(\ref{thermalpower3d}) is the rate equation of the LO phonon number.
Using the Fermi golden rule, it can be written for bulk systems~\cite{Tsai}
\[
\frac{dN_{\mathbf{q}}}{dt}\Bigr|_{c-LO}=2\frac{2\pi}{\hbar}|M_{\mathbf{q}}|^2\sum_{\mathbf{k}}I^2_{\mathbf{k},\mathbf{k}-\mathbf{q}}[(N_{\mathbf{q}}+1)f_{\mathbf{k}}(1-f_{\mathbf{k}-\mathbf{q}})
\]
\begin{equation}\label{KineticEquation3D}
	-N_{\mathbf{q}}f_{\mathbf{k}-\mathbf{q}}(1-f_\mathbf{k})]\delta(E_{\mathbf{k}}-E_{\mathbf{k}-\mathbf{q}}-\hbar\omega_{\mathbf{q}})
\end{equation}
where $M_{\mathbf{q}}$ is the electron-phonon scattering matrix element at wave vector $q$, $I^2_{\mathbf{k},\mathbf{k}-\mathbf{q}}$ represents the overlap of the wave functions of electron and phonon before and after the electron-LO phonon scattering (LO phonon emission or absorption process), $N_{\mathbf{q}}$ is the number of phonon at wave vector $\mathbf{q}$ and $f_{\mathbf{k}}$ is the Fermi-Dirac distribution of the electron at wave vector $\mathbf{k}$. The extra overall factor 2 in front of the right hand side counts for the spin degeneracy. It is emphasized that Eq.~(\ref{KineticEquation3D}) is valid only for 3D theory because the momentum conservation applies to all the three orthogonal directions during the electron-phonon scattering process.  
In the 3D theory derived in Ref.~\cite{Tsai}, the thermalization power finally takes the form of Eq.~(\ref{thermalpower3dbulk}) with the scattering rate given Eq.~(\ref{bulkelectronphononscatteringrate}).

In the presence of confinement, the above scattering rate has to be modified appropriately. The wave vectors for carriers will be noted ${\bf k}=({\bf k}_\perp, k_z)$. The longitudinal wave vector component $k_z$ takes discrete values of index $n$, that are related to the well width $L_z$~\cite{TannoudjiQM}.
Only transverse components of the wave vectors ($\mathbf{k}_{\perp},\mathbf{k}'_{\perp}=\mathbf{k}_{\perp}-\mathbf{q}_{\perp}$) satisfy momentum conservation. 
The rate of change of phonon number becomes
	\[
	\frac{dN_{\mathbf{q}_{\perp},q_z}}{dt}\Bigr|_{c-LO}=2\frac{2\pi}{\hbar}|M_{\mathbf{q}_{\perp},q_z}|^2\sum_{\mathbf{k}_{\perp}}\sum_{n,n'}I^2_{\mathbf{k}_{\perp},n;\mathbf{k}_{\perp}-\mathbf{q}_{\perp},n'}
	\]
	\[
	\times  \Bigl((N_{\mathbf{q}_{\perp},q_z}+1)f_{\mathbf{k}_{\perp},n}(1-f_{\mathbf{k}_{\perp}-\mathbf{q}_{\perp},n})
	-N_{\mathbf{q}_{\perp},q_z}f_{\mathbf{k}_{\perp}-\mathbf{q}_{\perp},n'}(1-f_{\mathbf{k}_{\perp},n})\Bigr)
	\]
	\begin{equation}\label{KineticEquation2D}   
		\times 
		\delta(E_{\mathbf{k}_{\perp},n}-E_{\mathbf{k}_{\perp}-\mathbf{q}_{\perp},n'}-\hbar\omega_{\mathbf{q}_{\perp},q_z})
	\end{equation}
%
where the electron-phonon scattering matrix element is given by~\cite{Tsai}
\begin{equation}\label{electronphononmatrixelement}
	|M_{\mathbf{q}}|^2=\frac{e^2\hbar \omega_{q_{\perp},q_z}}{2\varepsilon_0q^2V^{ph}}\left(\frac{1}{K_{\infty}}-\frac{1}{K_s}\right) \ ,
\end{equation}
with $q^2=(q_\perp^2+q_z^2)$ for the 2D formulation, $\omega_{q_{\perp},q_z} =\omega_{LO}$ as we have supposed LO phonons to be dispersionless, and $V^{ph}=S L^{ph}_z$ the volume of the LO phonon bath (not necessarily equal to the carrier volume) for which $S=L^2_{\perp}$ is the cross-sectional area of the device.  Moreover, $e$ is the electron charge, $\varepsilon_0$ the vacuum permittivity, $K_{\infty}$ the dielectric constant at infinite frequency and $K_s$ the static dielectric constant. 

\begin{figure}
	\includegraphics[width=\columnwidth]{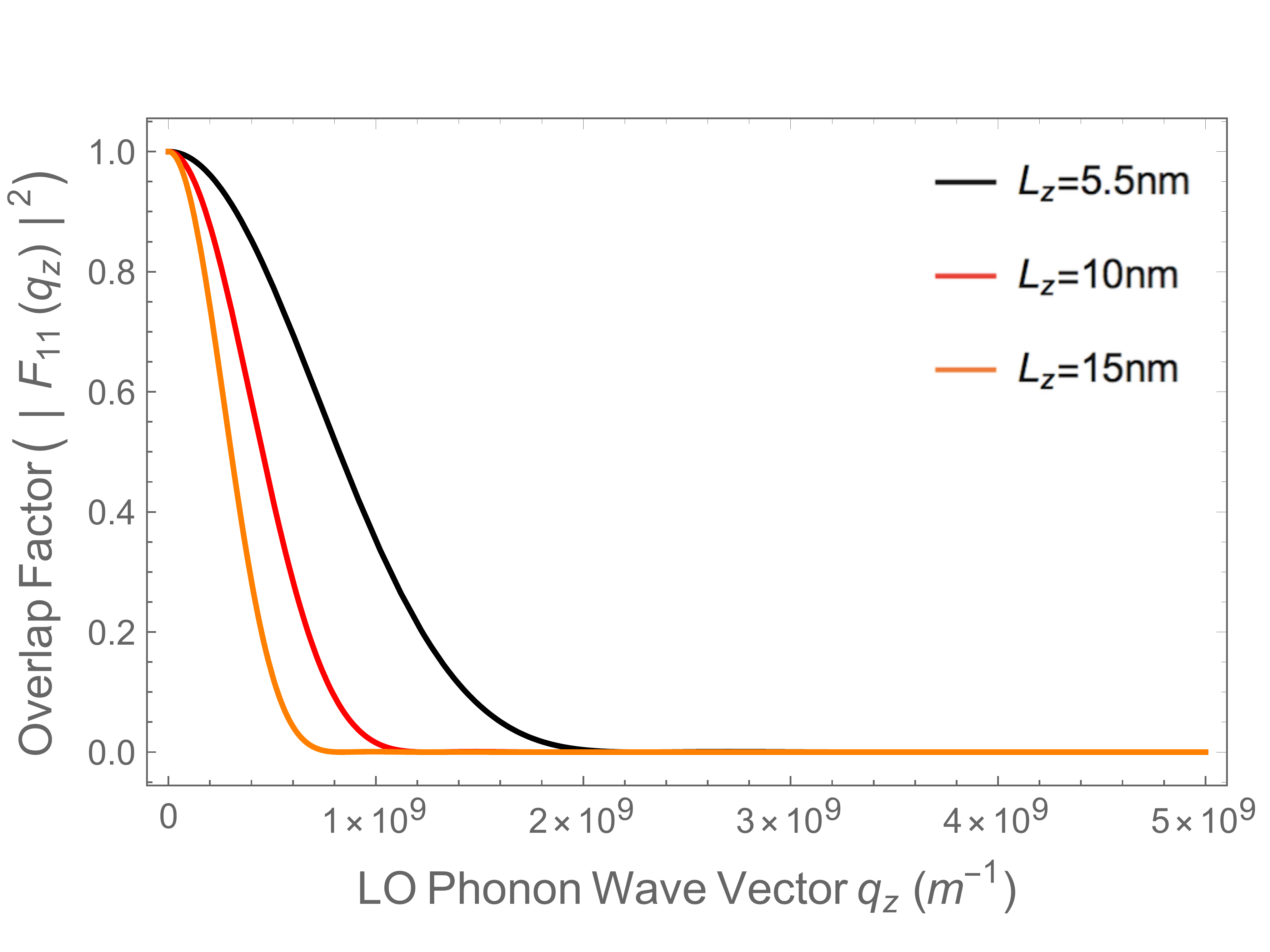}
	\caption{\label{fig:overlap} Profile of the overlap factor $|G_{11}(q_z)|^2$ from Eq.~(\ref{factor2}), for thicknesses $L_z=5.5$, $10$ and $15$ nm.}
\end{figure}

The wave function overlap $I^2_{\mathbf{k},\mathbf{k}-\mathbf{q}}$ occurring in Eq.~(\ref{KineticEquation2D}) is given in terms of the following factor~\cite{Price1981,dasSarmaMason}
\begin{equation}\label{wavefunctionoverlap} 
	I^2_{\mathbf{k}_{\perp},n,\mathbf{k}_{\perp}-\mathbf{q}_{\perp},n'}=|G_{n,n'}(q_z)|^2 \ ,
\end{equation}
where 
\begin{equation}\label{formfactor}
	G_{n,n'}(q_z)=\int^{L_z}_0 dz \psi_{n}(z)\exp(i q_z z)\psi^*_{n'}(z) \ ,
\end{equation}
with
\begin{equation}
	\psi_n(z)=\sqrt{\frac{2}{L_z}}\sin\left(\frac{n\pi z}{L_z}\right) \ ,
\end{equation}
relying on the approximation of infinite barrier height.
The calculation leads to
	\[
	G_{n,n'}(q_z)= \frac{i L_z q_z}{(n\pi - n' \pi + L_z q_z) (-n \pi + n' \pi + L_z q_z)} 
	\]
	\[
	+ \frac{
		i L_z q_z}{(n \pi + n' \pi - L_z q_z) (n \pi + n' \pi + 
		L_z q_z)}
	\]
	\[
	+ \frac{-i L_z q_ze^{i L_z q_z} \cos[(n - n') \pi] }{(n \pi - n' \pi + L_z q_z) (-n \pi + n' \pi + L_z q_z)} 
	\]
	\begin{equation}\label{factor2}
		-\frac{i L_z q_z e^{i L_z q_z} \cos[(n + n') \pi]}{(n \pi + n' \pi - 
			L_z q_z) (n \pi + n' \pi + L_z q_z)}\, .
	\end{equation}
The profile of $G_{n,n'}(q_z)$ is illustrated in Fig.~\ref{fig:overlap}, reflecting a distribution maximum at $q_z=0$ which sharpens with increasing the quantum well thickness. The overlap factor has a symmetry property $G_{n,n'}(q_z)=G_{n',n}(q_z)$.

\subsection{Intra-subband and inter-subband scattering rates}
\label{sec:intrainter}

The next step is the derivation of the electron-phonon scattering rate $dN_{\mathbf{q}_{\perp},q_z}/dt$, which was done with the hypothesis that thermalization occurs only in the conduction band, due to the large effective mass values of holes inside the valence band. Such an hypothesis helps to decipher the thermalization power behavior, but it remains easy to include all band contributions.

We evaluate the expression for the rate of change of the number of LO phonons Eq.~(\ref{KineticEquation2D}) using the electron-phonon scattering matrix element Eq.~(\ref{electronphononmatrixelement}) and the overlap factor Eq.~(\ref{wavefunctionoverlap}). The details of the derivation are given in Supp. Mat. 
Defining handy notations, for the scattering time, $\tau^{c-LO}_{\mathbf{q}_\perp,q_z}$, and its inverse the scattering rate $R_{\mathbf{q}_{\perp}, q_z}$, it can be written as
	\[
	\frac{dN_{\mathbf{q}_{\perp},q_z}}{dt}\Bigr|_{c-LO}
	=\frac{N_{\mathbf{q}}(T_c) - N_{\mathbf{q}_{\perp}, q_z}}{\tau^{c-LO}_{\mathbf{q}_\perp,q_z}}
	\]
	\begin{equation}
		=R_{\mathbf{q}_{\perp}, q_z}(N_{\mathbf{q}}(T_c) - N_{\mathbf{q}_{\perp}, q_z})
	\end{equation}
	where
	\[
	\frac{1}{\tau^{c-LO}_{\mathbf{q}_\perp,q_z}}=
	\frac{m_ce^2\omega_{LO}}{\pi\varepsilon_0 \hbar^2 q_{\perp}}\left[\frac{1}{K_{\infty}}-\frac{1}{K_s}\right]
	\frac{1}{(q^2_{\perp}+q^2_z)L^{\mathrm{phon}}_z} \ \ \times 
	\]
	\[
	\sum_{n,n'}|G_{n,n'}(q_z)|^2\int^{+\infty}_{k^{\mathrm{min}}_{\perp}(n,n')} dk_{\perp}
	\frac{k_{\perp}}{\sqrt{k^2_{\perp}-(k^{\mathrm{min}}_{\perp}(n,n'))^2}} 
	\]
	\begin{equation}\label{rateofchangeofbosonnumberbis}
		\times\left[f(E_{\mathbf{k}_{\perp},n}-\hbar \omega_{LO})-f(E_{\mathbf{k}_{\perp},n})\right] 
		\ .
	\end{equation}

We have defined
\begin{equation}\label{eq:Rnnp}
	k^{\mathrm{min}}_{\perp}(n,n')= \Bigl| \frac{q_{\perp}}{2}+\frac{m_c\omega_{LO}}{\hbar q_{\perp}} +  \frac{({n'}^2-n^2)}{2 q_{\perp}}\frac{\pi^2}{L^2_z}\Bigr|
\end{equation}
the minimum transverse wave vector for an electron to emit an LO phonon. 

Upon emitting a phonon of energy $\hbar\omega_{LO}$, an electron may stay in the same subband, which constitutes an intra-subband scattering process, or move to another subband, which represents an inter-subband scattering process. 
Splitting the intra-subband and inter-subband contributions,
the LO phonon emission rate for each phonon wave vector $\mathbf{q}=(\mathbf{q}_{\perp},q_z)$ can be written as
\begin{equation}\label{cLOscatteringrate}
	\frac{1}{\tau^{c-LO}_{\mathbf{q}_{\perp}, q_z}}=\sum^{N_s}_{n=1}\frac{1}{\tau^{c-LO}_{\mathbf{q}_{\perp}, q_z}(n,n)}+\sum^{N_s}_{n,n'=1(n\neq n')}\frac{1}{\tau^{c-LO}_{\mathbf{q}_{\perp}, q_z}(n,n')},\ 
\end{equation}
with $N_s$ the number of subbands inside the well.
We have thus introduced $1/\tau^{c-LO}_{{\bf q}_\perp,qz} (n,n')$ as the net phonon emission rate  from subband $n$ to subband $n'$. It is not symmetric upon $n-n'$ permutation.
It is important to note that our theory only takes into account the scattering between electron states confined in the quantum well and the LO phonons. That is, we do not consider scattering processes involving electron states whose energy are higher than the InGaAsP barrier height in Fig.~\ref{fig:fig2}.  

\begin{figure*}[ht!]
	\includegraphics[width=8.5cm]{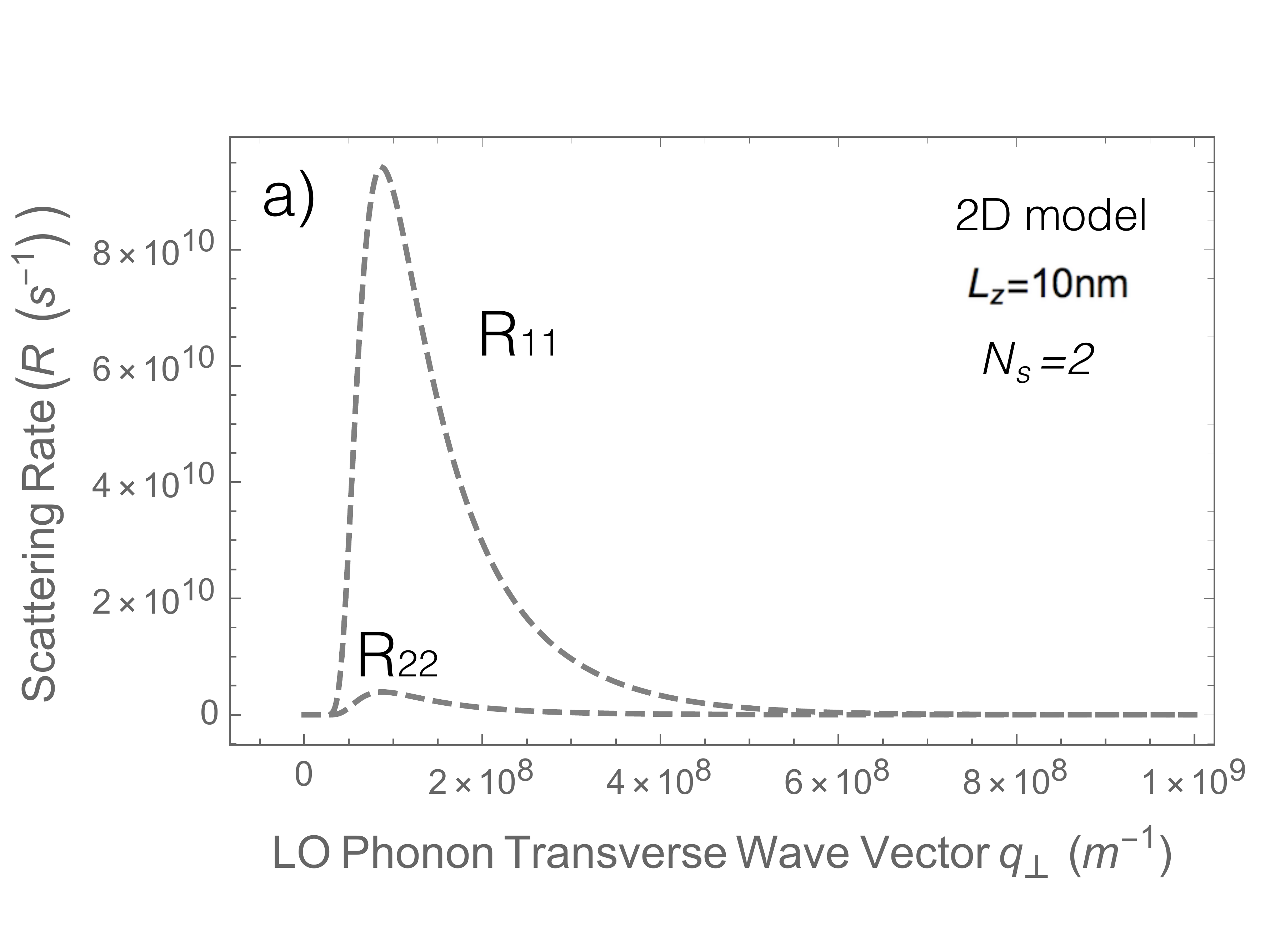}
	\includegraphics[width=8.5cm]{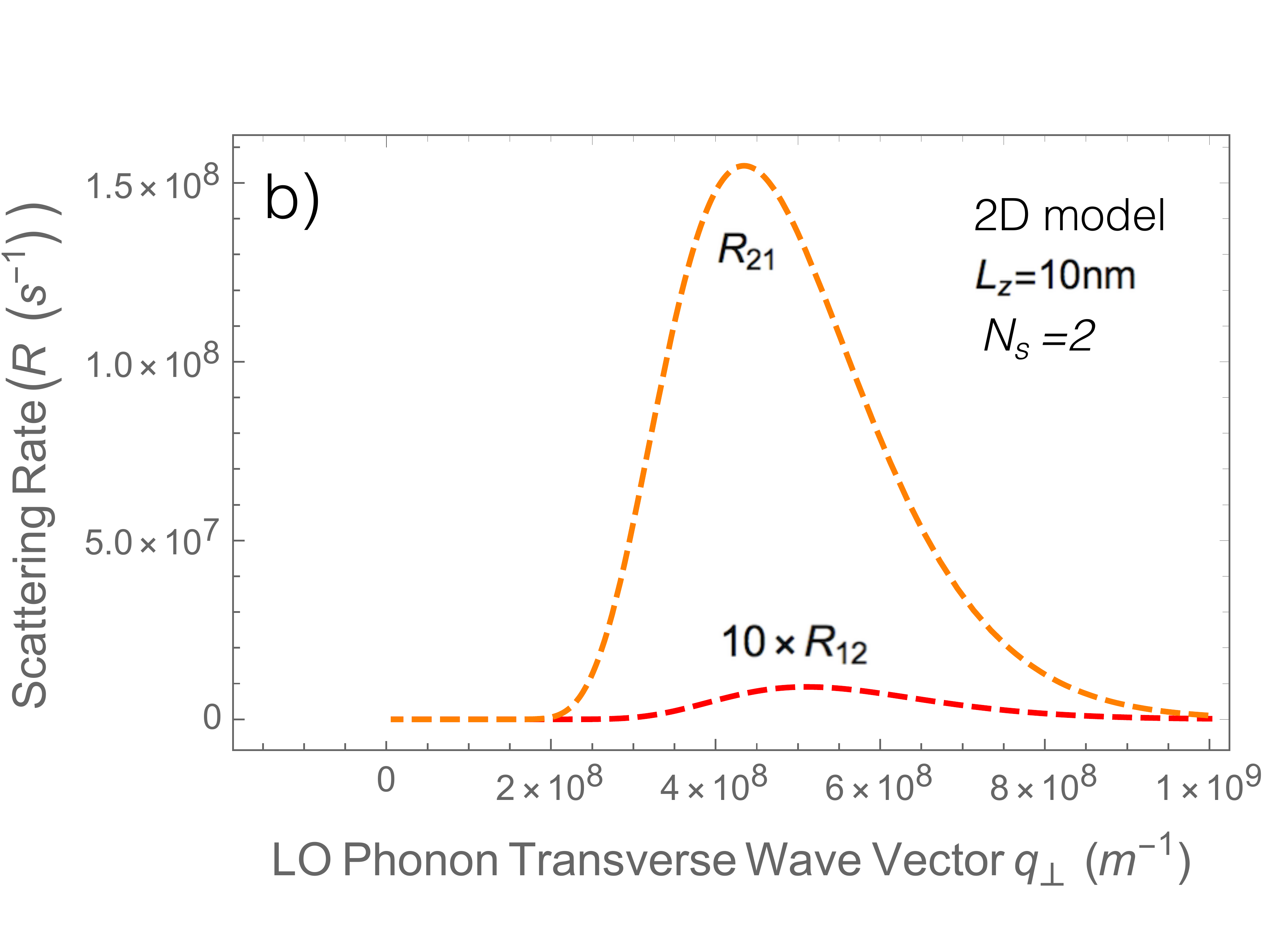}
	\includegraphics[width=8.5cm]{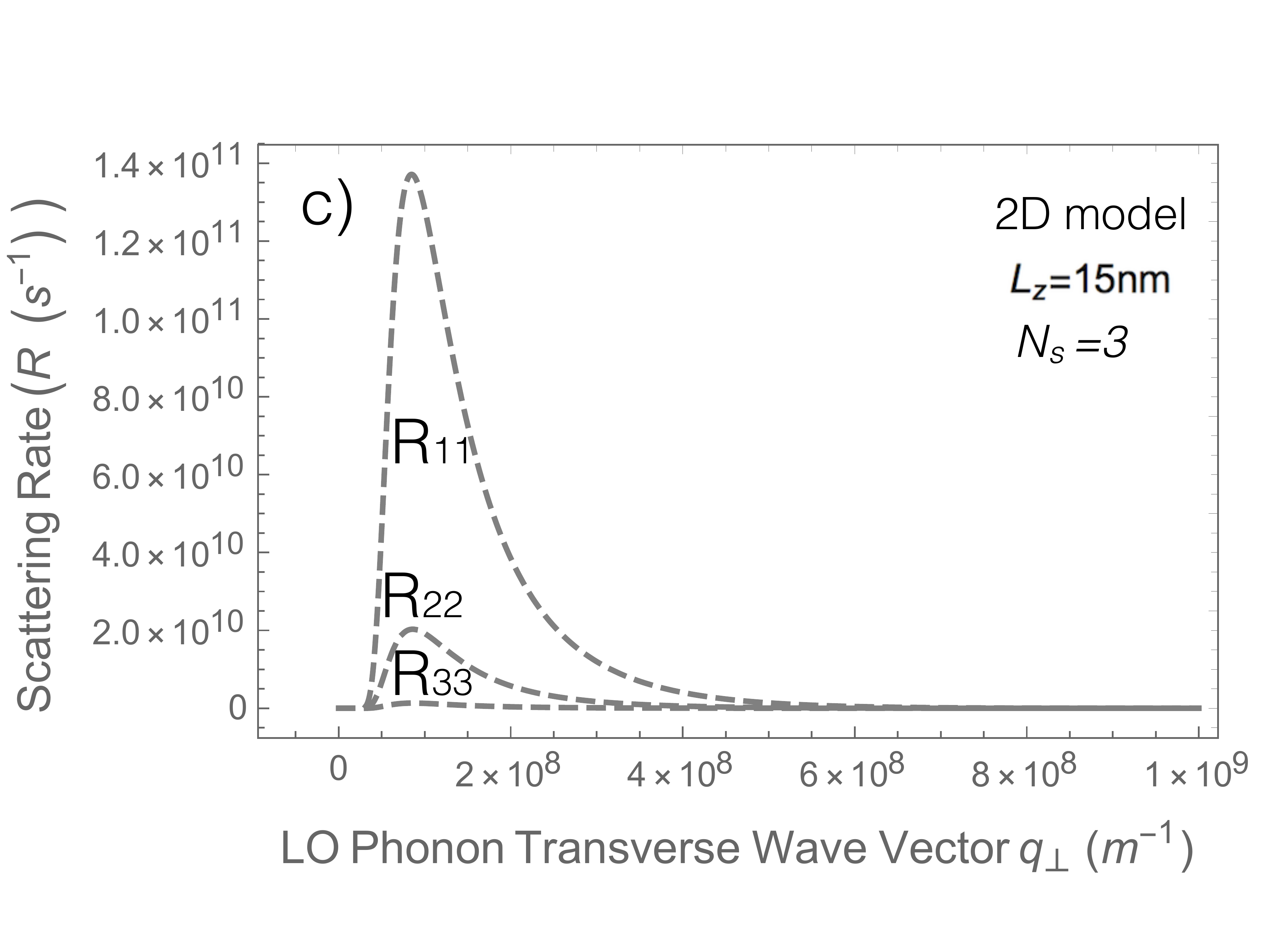}
	\includegraphics[width=8.5cm]{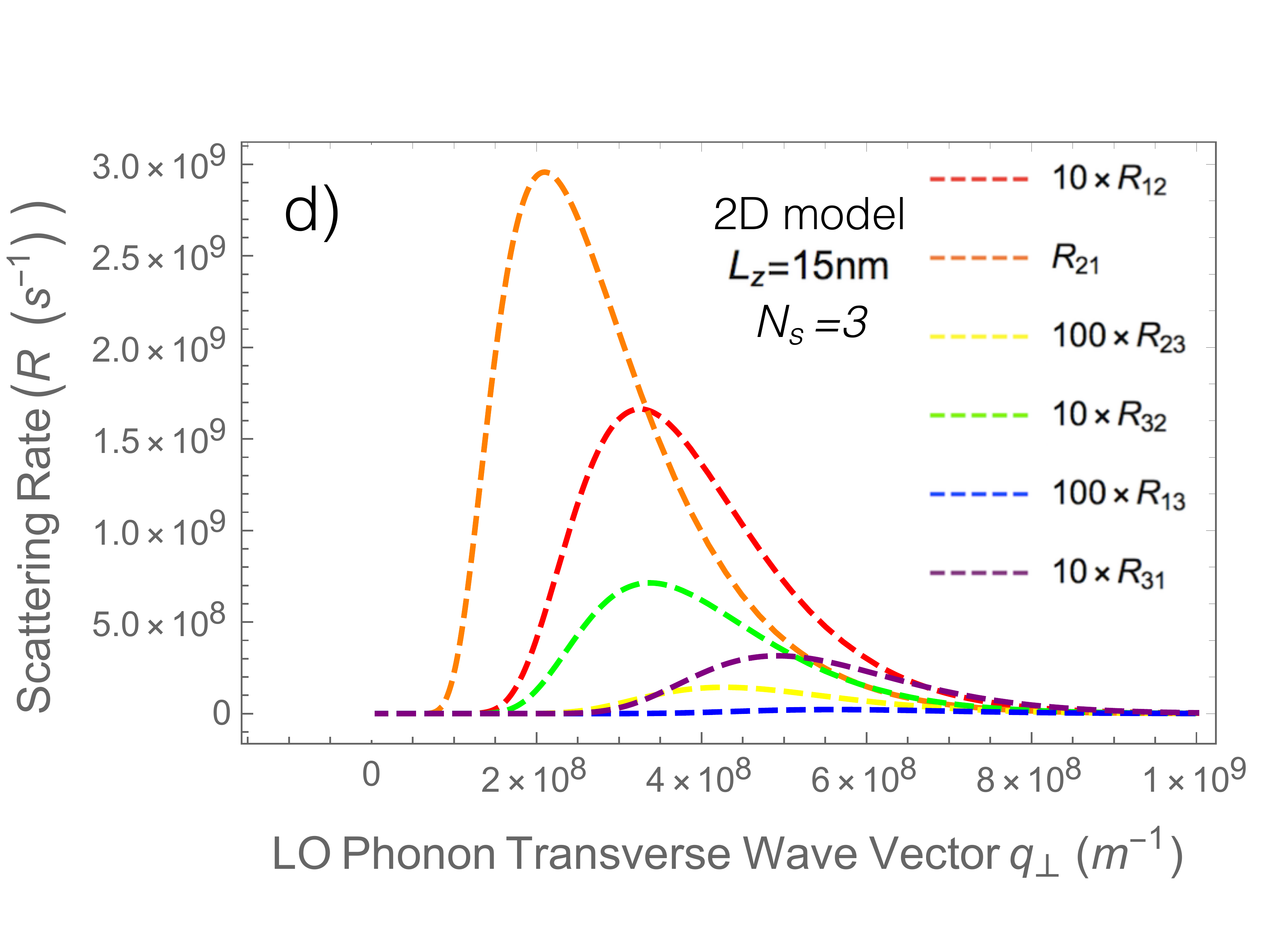}
	\caption{\label{fig:fig4}   Intra (grey curves) and inter-subband (colored curves) scattering rates $R_{n,n'}$ (see Eq.~(\ref{eq:Rnnp}) for definition) in 10~nm a) and b), and 15~nm c) and d) quantum wells, as functions of $q_{\perp}$ for $T_c=450$K and $T_L=300$K. Other parameters are $m_c=0.041m_0$ with $m_0$ the bare electron mass, $\hbar\omega_{LO}=36$meV, $K_{\infty}=10.9$, $K_s=12.9$, while the chemical potential $\mu_c=-35.9074$~meV for $10$~nm and $\mu_c=-37.7681$~meV for $15$~nm quantum wells.}
\end{figure*}

The numerical results for the summed scattering rates 
\begin{equation}
	R_{nn'} = L^{ph}_z \int  \frac{dq_z}{2 \pi} \frac{1}{\tau^{c-LO}_{{\bf q}_\perp,qz} (n,n')} 
\end{equation} 
are shown in Fig.~\ref{fig:fig4} both for intra-subband $(n'=n)$ and inter-subband $(n' \neq n)$ processes.
We observe that he intra-subband contribution of the first subband, $R_{11}$,  gives the largest contribution to the electron-phonon scattering rate, and that intra-subband contributions of higher subbands decrease with the subband index, a conclusion which is in agreement with that of the classic work by Riddoch-Ridley~\cite{Ridley2}.  Moreover, for small thicknesses where only few subbands exist in the well  ($N_s=2$ for $L_z=10$~nm and $N_s=3$ for $L_z=15$~nm), the $R_{21}$ inter-subband term is much larger than other inter-subband contributions to the scattering rate. Eventually, as the thickness is increased further, the number of subbands increases, and the inter-subband scattering rate (summing over all inter-subband processes) overtakes that of the first intra-subband scattering rate. 

\subsection{Impact of non-equilibrium phonon distribution hypothesis}
\label{sec:noneqphonon}

A key mechanism that is believed to drive hot carrier effect is the occurrence of a non-equilibrium distribution of LO phonons. This hypothesis means that the LO phonon occupation number do not match a Bose-Einstein function defined at the lattice temperature $T_L$ with which acoustic phonons stay in equilibrium. The effect of a non-equilibrium LO phonon distribution seems to be significant in bulk 3D systems~\cite{Tsai}, but this is not necessarily the case in quantum well heterostructures, due to the quantum confinement which constrains the interaction between electrons and phonons.  

\begin{figure*}[ht!]
	\includegraphics[width=8cm]{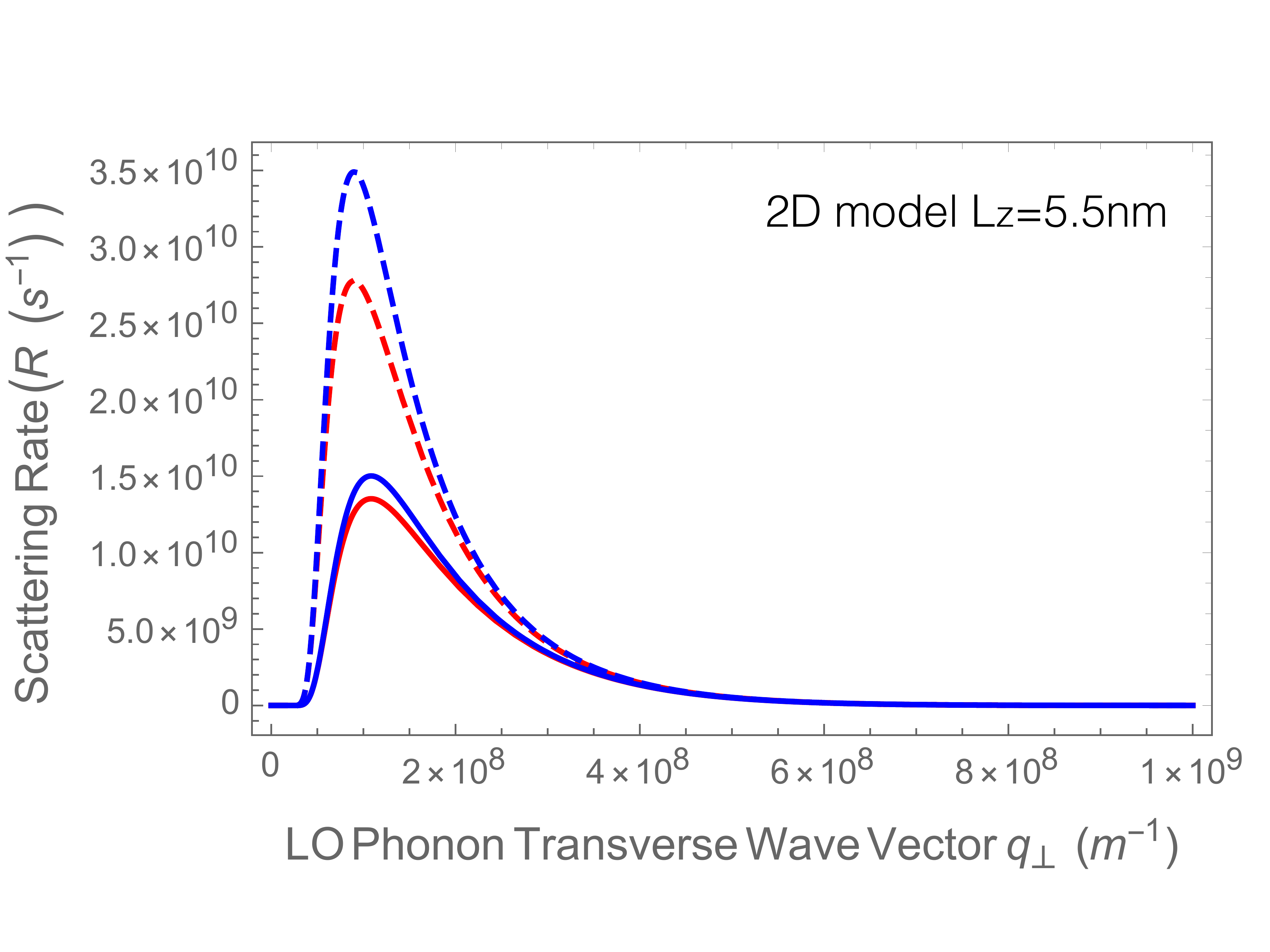}
	\includegraphics[width=8cm]{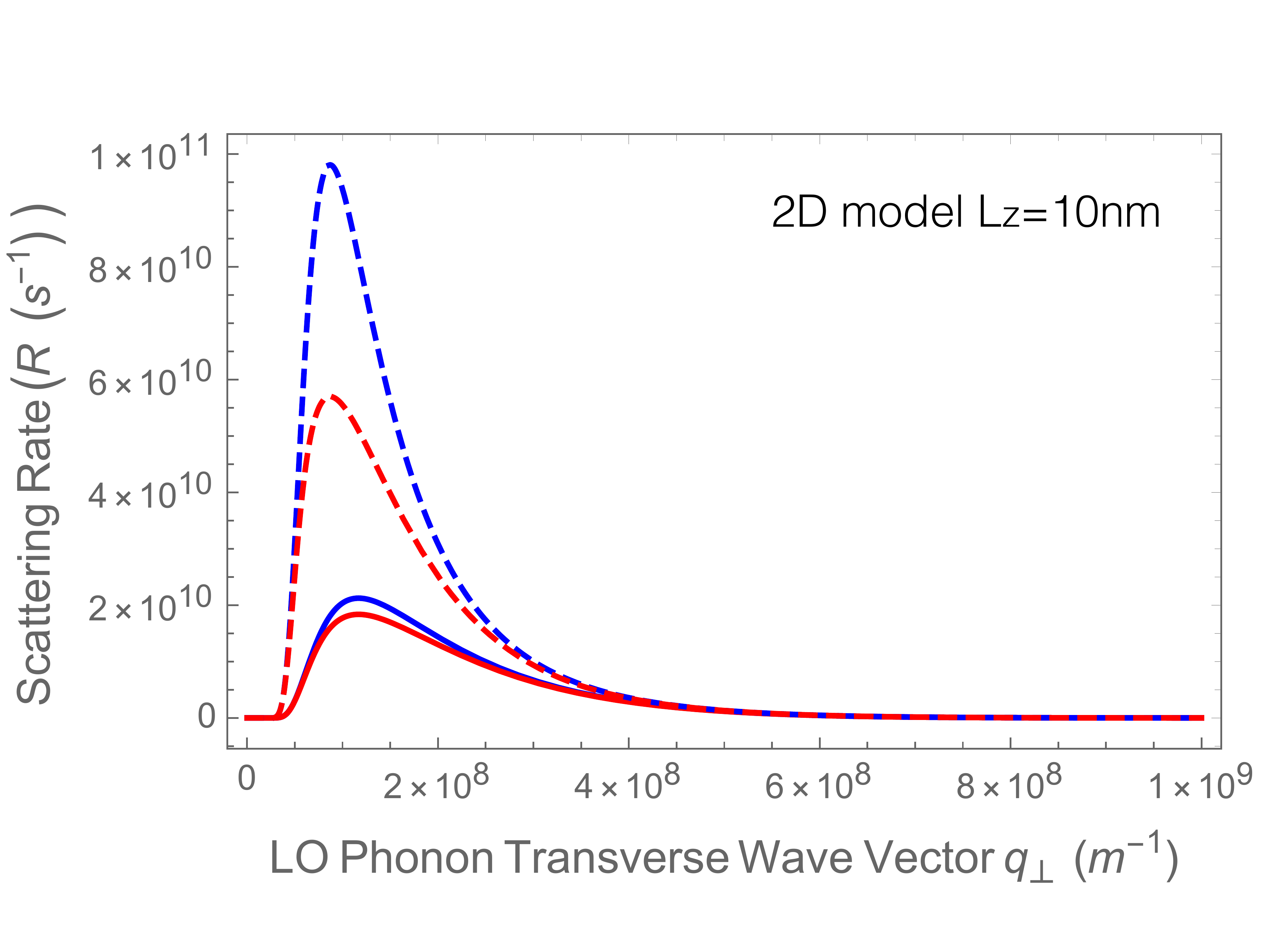}
	\includegraphics[width=8cm]{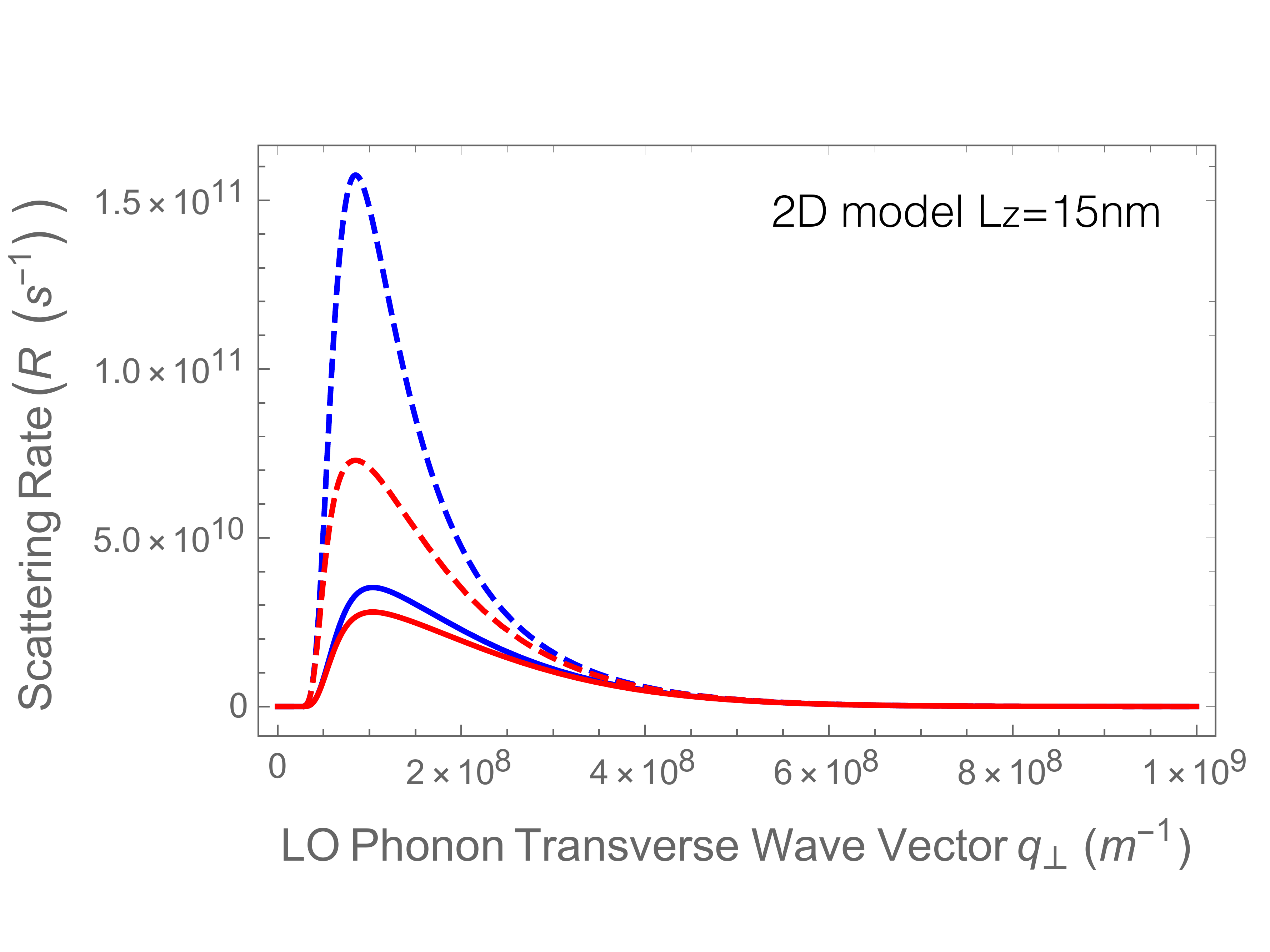}
	\includegraphics[width=8cm]{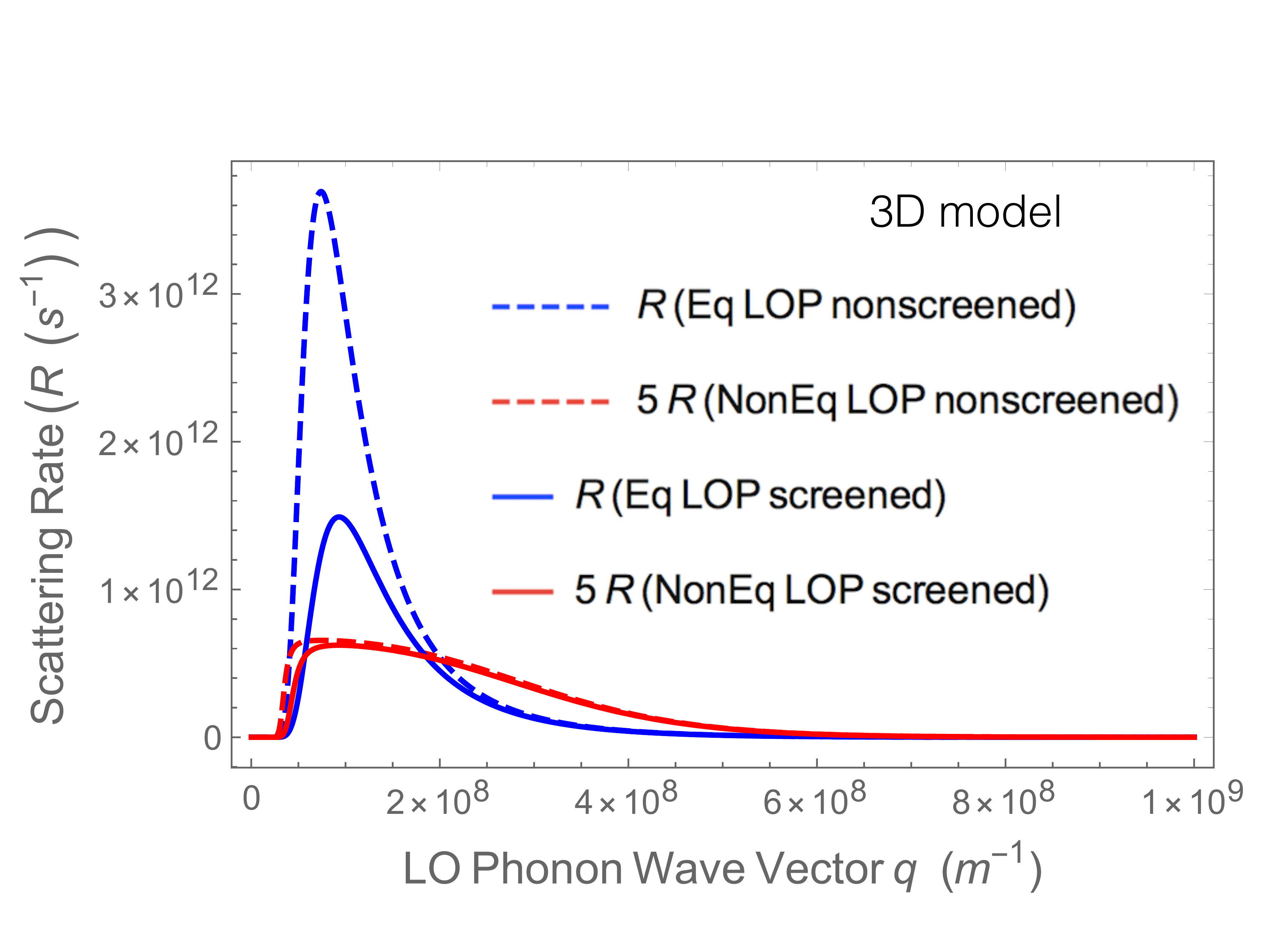}
	\caption{\label{fig:fig5} Electron-LO phonon scattering rates as defined in Eq.~(\ref{eq:Neqphonon}) for a non-equilibrium (NonEq, red curves) and an equilibrium (Eq, blue curves) LO phonon distribution, as functions of  $q_{\perp}$.
		Solid and dashed lines correspond to the cases with and without screening respectively.
		Results are shown for three quantum wells of  thicknesses $L_z=5.5$, $10$, and $15$nm, as well as for bulk 3D model. For the sake of clarity, 3D scattering rates for non-equilibrium phonons have been multiplied by a factor 5 (lower right quadrant).
		The parameter values are the same as in Fig. \ref{fig:fig4}. }
\end{figure*}

In this section, we therefore investigate the relative strengths of the effects of confinement and non-equilibrium phonon distribution on the electron-LO phonon scattering rate. 
With LO phonons characterized by an a-thermal distribution specified in Eq.~(\ref{LOphononpopulationnumber}), the thermalization power is given by Eq.~(\ref{thermalpower3dbulk}), while, with LO phonons at the lattice temperature $T_L$, henceforth called equilibrium phonons,  the thermalization power is given by Eq.~(\ref{thermalpower3dequilibriumLOphonons}). 
To this end, using Eqs.~(\ref{rateofchangeofbosonnumberbis}) and (\ref{eq:tauLO-ac}), we plot in Fig.~\ref{fig:fig5}  the two integrated scattering rates
\begin{equation}\label{eq:Neqphonon}
	R^{\rm{NonEq}(\rm{Eq})}({\bf{q}_\perp})=L^{ph}_z \int \frac{d q_z}{2 \pi} \frac{1}{\tau^{c-LO}_{\bf{q}_\perp, q_z}+(\tau^{LO-ac}_{\bf q})}\, ,
\end{equation}
where $\tau^{LO-ac}_{\bf q}$ in parenthesis is present or absent respectively in the non-equilibrium (NonEq) and equilibrium (Eq) cases. 
With the same convention, the 3D results, $1/(\tau_{{\bf q}}^{c-LO} +(\tau^{LO-ac}_{\bf q}))$ from Eqs.~(\ref{bulkelectronphononscatteringrate}) and (\ref{eq:tauLO-ac}) are also represented.
In the figure, the solid curves encompass a screening effect that will be discussed in the following section. 
Focussing on the dashed curves of Fig.~\ref{fig:fig5}, the results show that the expected reduction of the scattering rate due the non-equilibrium phonon distribution hypothesis is suppressed for thinner wells, while it remains strong in the 3D case.

The ratio $\tau^{LO-ac}_{\bf q} / \tau_{{\bf q}}^{c-LO}$ provides a simple tool to explain why the LO phonons have a weaker tendency to go out-of-equilibrium in quantum wells.
On the one hand, the scattering rate $1/\tau^{c-LO}_{\mathbf{q}}$ is strongly reduced due to confinement relatively to bulk, as shown in section~\ref{sec:intrainter}.
On the other hand, the LO phonon-acoustic phonon scattering rate remains unchanged, as phonons are assumed to be unconfined, $1/\tau^{LO-ac}_{\mathbf{q}} \simeq 1.35\, 10^{11}$~s$^{-1}$.
Starting from the 3D case, the characteristic times are in ratio $\tau^{LO-ac}_{\bf q} / \tau_{{\bf q}}^{c-LO} \simeq 30$ for the wave vector corresponding to the maximum of the scattering rate, the corresponding phonon number $N_{\bf q}$ from Eq.(\ref{LOphononpopulationnumber}) is then leaning towards $N_{\bf q}(T_c)$: the a-thermal distribution of LO phonons approaches a hot equilibrium distribution. 
In contrast, in the quasi 2D case, and more particularly for narrow quantum wells, 
the scattering rate $1/ \tau^{c-LO}_{\mathbf{q}}$ is strongly reduced due to confinement, the reduction can be as large as 2 orders of magnitude examining Fig.~\ref{fig:fig5}.
The ratio $\tau^{LO-ac}_{\bf q} / \tau_{{\bf q}}^{c-LO}$ thus becomes smaller than 1, which steers the corresponding phonon number towards $N_{\bf q}(T_L)$: the a-thermal distribution of LO phonons approaches the lattice temperature equilibrium distribution of acoustic phonons. 
This analysis may be interpreted as follows: in quantum well solar cells, that once an LO phonon is emitted by an electron, it never gets re-absorbed back by the electron. Instead, the LO phonon quickly transforms into acoustic phonons which stay in equilibrium at the  lattice temperature. In other words, LO phonons quickly cool down to the lattice temperature, suppressing the non-equilibrium LO phonon effect. Hot carrier effect is thus controlled by confinement in reduced dimensionality.

\subsection{Dependence on Electron-Electron Screening Effect}
\label{sec:screening}

So far, we did not consider the screening of the electron-phonon interaction in Eq.(\ref{electronphononmatrixelement}). However, screening ultimately renormalizes  the scattering matrix element $|M_{\mathbf{q}_{\perp},q_z}|^2$ of Eq. (\ref{KineticEquation2D}) as
\begin{equation}
	|M^{\mathrm{screen}}_{\mathbf{q}_{\perp},q_z}|^2=|M_{\mathbf{q}_{\perp},q_z}|^2\left(\frac{\varepsilon_0}{\varepsilon_{\mathrm{screen}}(\mathbf{q}_{\perp},q_z,\omega_{LO})}\right)^2.
\end{equation}
Making a static and strictly 2D approximation,  we have to evaluate the following screening function~\cite{RidleyBook2}
	\[
	(\varepsilon_{\mathrm{screen}})^{n_4,n_3}_{n_2,n_1}(\mathbf{q}_\perp,0)=\varepsilon_0\delta_{n_2,n_4}\delta_{n_1,n_3}
	\]
	\begin{equation}\label{screeningfunction}
		-\frac{ e^2}{2q_{\perp}S}
		F^{n_4n_3}_{n_2n_1}(\mathbf{q}_\perp)
		\sum_{{\mathbf{k}_\perp}, \sigma} \frac{f(E_{n_2,\mathbf{k}_\perp+\mathbf{q}_\perp})-f(E_{n_1,\mathbf{k}_\perp})}{E_{n_2,\mathbf{k}+\mathbf{q}_\perp}-E_{n_1,\mathbf{k}_\perp}}
	\end{equation}
where $F^{n_4,n_3}_{n_2,n_1}(\mathbf{q})$ is the form factor of the electron-electron interaction involving four indices for the confined electron subbands $n_1,n_2,n_3,n_4$. $\sum_\sigma$ indicates the sum over the spin.

The number of form factor tensor elements involved in the calculation increases with the number of subbands, going as $N^4_s$ where $N_s$ is the number of subbands. 
For the sake of simplicity, we make the approximation of accounting for only the screening of the first intra-subband process, with
\begin{equation}\label{Ffunction}
	F^{11}_{11}=\frac{1}{2} \frac{\eta (\eta^2 + \pi^2) (3 \eta^2 + 2 \pi^2) - 
		\pi^4 (1 - e^{-2 \eta})}{[\eta (\eta^2 + \pi^2)]^2}
\end{equation}
where $\eta=q_{\perp}L_z/2$~\cite{RidleyBook2}. We thus ignore the screening of all other scattering processes. This is justified by the fact that the first intra-subband scattering rate $R_{11}$ strongly dominates over all other scattering rates $R_{nn'}$, as can be seen from Fig.~\ref{fig:fig4}.
As a consequence, the present approximation which underestimates the screening effect, overestimates most scattering rates.

Figure~\ref{fig:fig5} gathers the effects of screening on the electron-phonon scattering rates in the bulk and in quantum wells, for LO phonons out-of equilibrium or at equilibrium with the lattice. 
As a general trend, screening renormalizes downward the scattering rate, whatever the hypothesis made for the LO phonon distribution.
The reduction is more pronounced for wider quantum wells up to the 3D limit. 
However, at the smallest thickness where the confinement effect is strongest, it appears that the hypothesis of out-of equilibrium LO phonons has only a small influence on scattering rates due to an increasing impact of $\tau^{LO-ac}_{\bf q}$, whatever screening is included, which complies with the analysis we have developed section~\ref{sec:noneqphonon}.


\section{Signatures of Quantum Confinement and Enhancement of Hot Carrier Effect in terms of Thermalization Power}

\subsection{Calculation protocol}

The objective of this section is to discuss the power needed to maintain a carrier population at fixed temperature $T_c$ and quasi-Fermi level splitting $\Delta \mu$~\cite{PWurfel}, when confinement increases in InGaAsP/InGaAs/InGaAsP quantum well heterostructures shown Fig.~\ref{fig:fig2}.
These two thermodynamic parameters have been extracted from photoluminescence experiments done for a particular quantum well, which makes the obtained conclusions more reasonable and exploitable to interpret real situations. We thus have $T_c=450$~K and  $\Delta \mu= 0.653$~eV, and we keep it constants all along the study.
For each quantum well thickness, these values are used to numerically estimate the chemical potential $\mu_c$ from the electroneutrality equation.
Details of the calculation protocol are given in {Supp. Mat}.
For LO phonons, we take $\hbar\omega_{\mathbf{q}}\equiv\hbar\omega_{\mathbf{q}_{\perp},m}=\hbar\omega_{LO}=36$~meV in the calculations, independent of wave vector. Additionnally, acoustic phonons are assumed to be in an equilibrium distribution at the lattice temperature $T_L=300$~K.
The 2D thermalization power is finally computed within our theoretical model combining Eqs.~(\ref{rateofchangeofbosonnumberbis}), (\ref{thermalpower2d}) and (\ref{eq:tauLO-ac}). 

\subsection{Findings: 2D thermalization power}\label{sec:findings}

\begin{figure*}
		\centering
		\includegraphics[width=16cm]{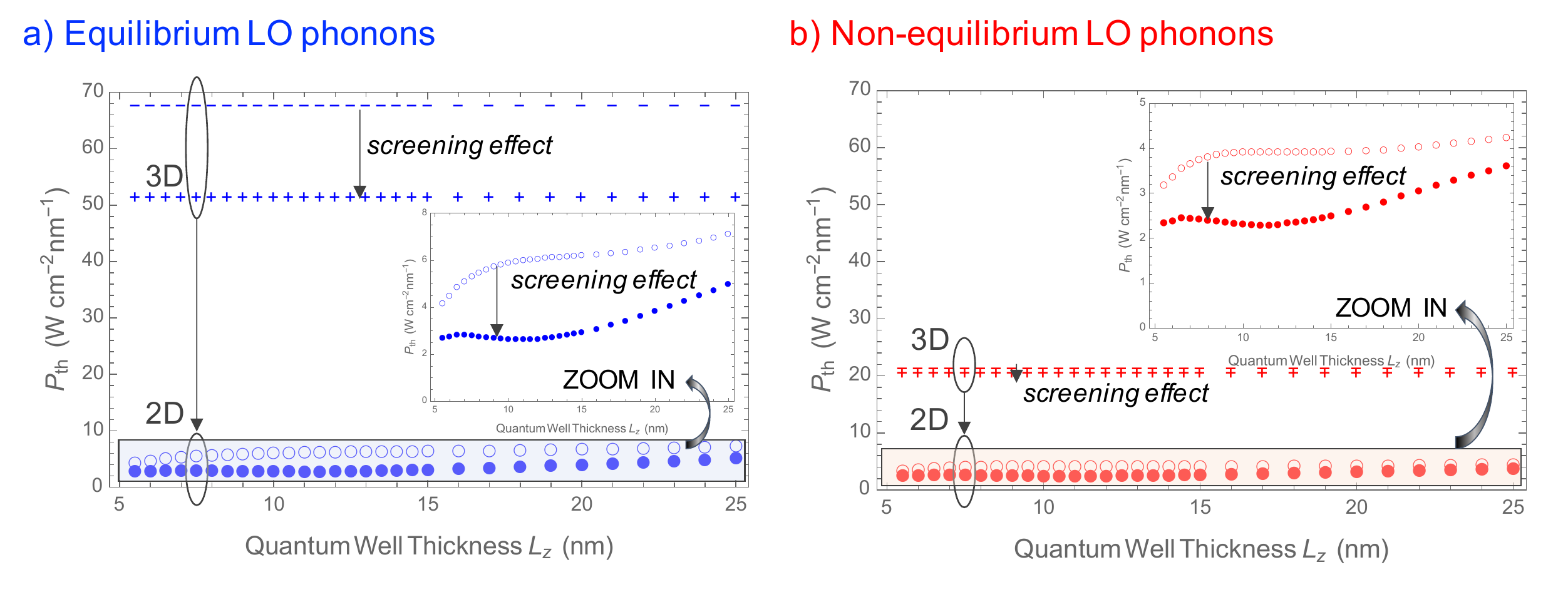}
		\caption{Thermalization power per unit volume as a function of the quantum well thickness in InGaAs heterostructure, considering a) an equilibrium (in blue) b) a non-equilibrium (in red) LO phonon distribution, with (filled circles) and without (empty circles) screening effect of electrons computed from the 2D model using Eq.~(\ref{thermalpower2d}), and from the 3D model using Eq.~(\ref{thermalpower3dbulk}). Input parameters are $T_c=450$~K, $\Delta \mu= 0.653$~eV and $T_L=300$~K. }\label{fig:2DThermPower}
	\end{figure*}

We compare the  thermalization power resulting from the proposed 2D theoretical model with the one obtained from the 3D model of Ref.~\cite{Tsai}, within different frameworks of hypothesis shown Fig.~\ref{fig:2DThermPower}: a) without (in blue) and b) with (in red) the assumption that LO phonon distribution is non-equilibrium following the derivations of section~\ref{sec:noneqphonon}, and including or not the screening effects as described in section~\ref{sec:screening}.

The first important observation that can be made is that the thermalization power density is lower in 2D than in 3D systems, whatever the used assumptions. It means that the amount of power needed to achieve a hot carrier population at given temperature and Fermi level splitting is lower in quantum well heterostructures than in the bulk material.
The result suggests that reducing the system's dimensionality reduces the system's propensity for thermalization.
Confinement thus enhances the hot carrier effect,  which is expected  to make quantum well based solar cells more efficient, due to the fact that the quantification of electronic states increases  the amount of carrier energy to be converted into electrical power because it reduces the thermalization power.
Actually, the reduction of the scattering rate  in quantum wells was also pointed out in more sophisticated models of electron-LO phonon interaction~\cite{Lassnig, RiddochRidley, Sawaki}. Moreover, the thermalization power reduction of about one order of magnitude is consistent with existing works in the literature, e.g.~\cite{MonteCarlo}.

In the 3D bulk case, the thermalization power density is constant, as can be seen in Fig.~\ref{fig:2DThermPower}.
This behavior is in agreement with experimental results where the thermalization power is found to vary linearly with the thickness of the absorber for relatively thick samples of 100-200 nm, where the physics approaches well that of a bulk system~\cite{GiteauJAP}.
In contrast, the 2D thermalization power density increases with the quantum well thickness, but remains always lower than the value computed in the 3D model under the same conditions: LO phonons in equilibrium or not, with screening effect taken into account or not.
As a general trend, the assumption of a non-equilibrium LO phonon distribution always reduces the thermalization power, by an amount decreasing with the quantum well thickness for the 2D model whereas it is independent of thickness for the 3D model.

\begin{figure}
	\centering
	\includegraphics[width=8cm]{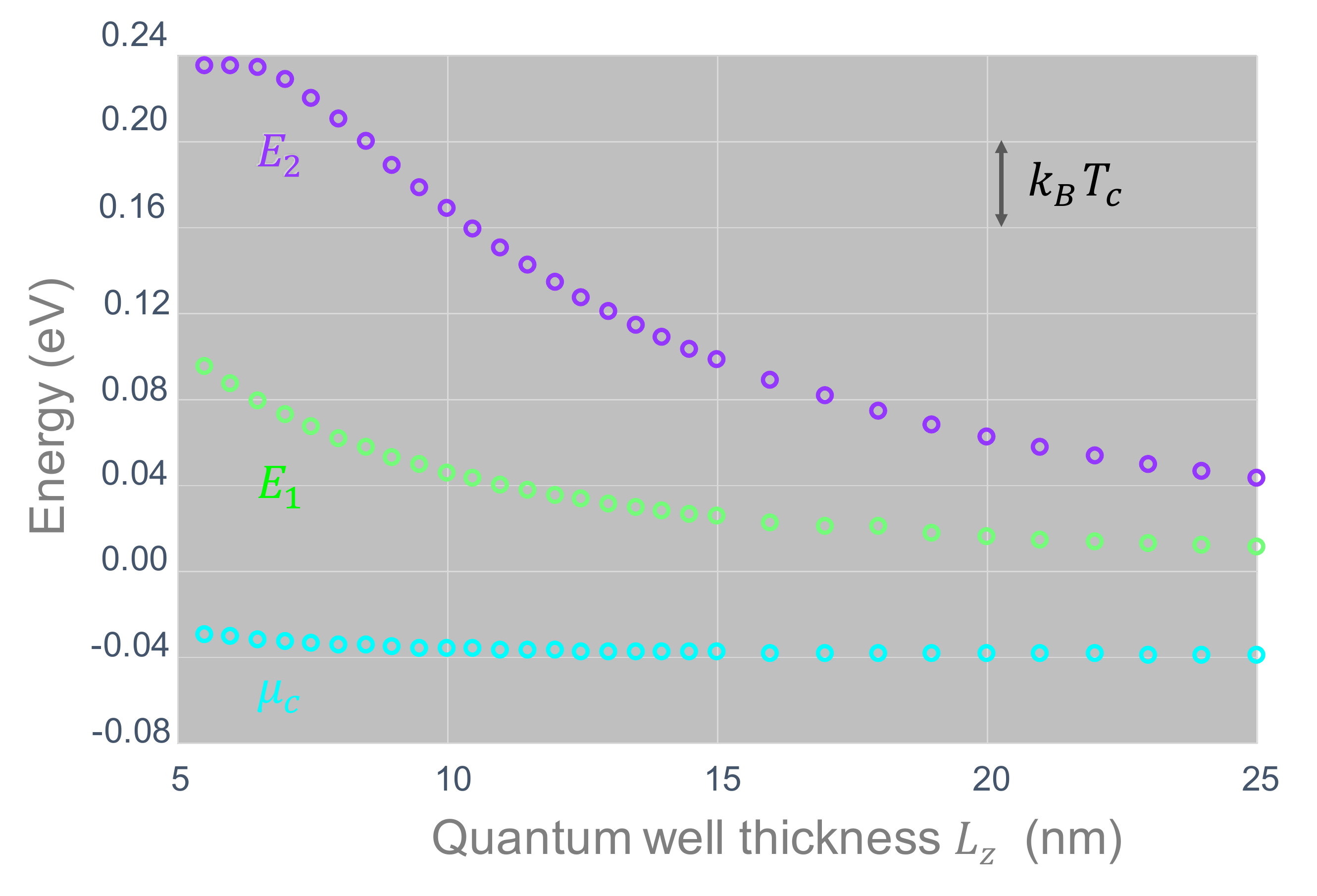}
	\caption{First and second conduction subband minima and quasi Fermi level, as a function of the quantum well thickness. The energy origin is at the bulk conduction band minimum.Thermal energy, $k_BT=0.039$~eV, is also represented.}\label{fig:fillings}
\end{figure}

Moreover, a non-trivial dependence on the quantum well thickness clearly appears in the insets of Figs.~\ref{fig:2DThermPower} a) and b).
Ignoring screening (empty circles), the thermalization power exhibits a staircase shape with a quasi-plateau for thicknesses between 8 and 16~nm, which reflects the evolution of subband filling as the quantum well thickness increases.
At small well thicknesses, each subband starts to fill significantly when the energy separation between the subband minimum and the Fermi level goes bellow $4k_BT_c\approx 4\times39$~meV (at which Fermi occupation goes beyond 0.01). This evolution is represented for the two first subbands in Fig.~\ref{fig:fillings}. As the well thickness increases, the filling increases. However in 2D systems, the density of states is proportional to $1/L_z$ inside each subband (for a fixed  Fermi energy), so the carrier density must finally decrease at larger well thickness.
The thermalization increase by plateau is thus  interpreted by distinguishing the intra- and inter-subband contributions to the 2D thermalization power of the first and second subbands, as represented Fig.~\ref{fig:subbandpower}. 
\begin{figure}
	\centering
	\includegraphics[width=8cm]{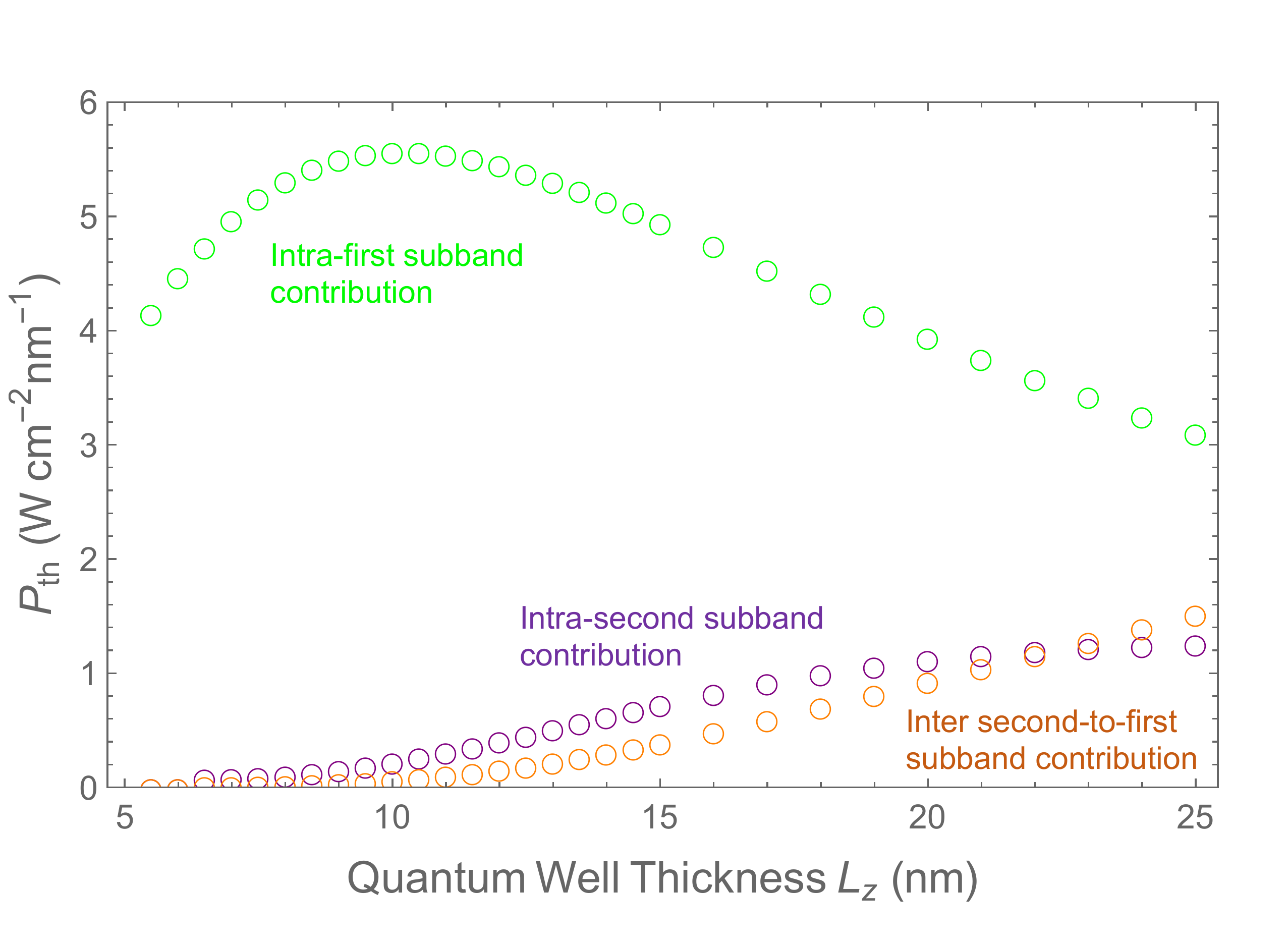}
	\caption{Detailed subband contributions to the 2D thermalization power per unit volume as a function of the quantum well thickness, in the simplest framework of an equilibrium LO phonon distribution, without screening effect.}\label{fig:subbandpower}
\end{figure}
At first, the separation energy between the first subband minimum and the quasi Fermi level decreases below $4k_BT_c$, while the energy separation between the second subband and the quasi Fermi level remains above this value, as can be seen  Fig.~\ref{fig:fillings}. Consequently, the intraband contribution of the first subband increases then decreases when the $1/L_z$ term dominates. %
In a second stage, above 12~nm, the second subband approaches the quasi Fermi level $\mu_c$, it starts to fill and participates to thermalization while the first subband depopulates. The first intra-subband contribution to power decreases while both the inter- and intra- second subband contributions increase. 
In total, the thermalization power plotted Fig.~\ref{fig:2DThermPower} increases from smaller well thicknesses, then saturates when the contributions of two first subbands compensate, next starts to increase again when intra- and inter- second subband components increase faster than the intra- first subband component decreases.
In these confining heterostructures,  the thermalization process thus follows several channels that are activated one after the other. 
The occurence of a plateau transition depends on temperature, but also on confinement energies, that are directly related to material nature throughout effective mass and barrier potential.
The proposed analysis also explains why the 2D thermalization power remains far from the 3D value at 25~nm, because only two subbands are at play for this thickness value, while there are already five subbands inside the well at 25~nm.

Finally, screening in general also reduces thermalization power, by an amount that varies with thickness in the 2D model, while it is independent of thickness in the 3D model. 
In particular, the screening-induced reduction of power is non-monotonic in quantum well heterostructures.
As a result, the thermalization power density passes through a minimum in the plateau region, before increasing again as one keeps increasing the thickness of the quantum well layer. 
This non-monotonic behavior thus emerges as a signature of energy quantization in 2D systems.

\section{Comparison with Experiments}

Real systems employed in experiments are far more complex than what we model in the present study.
We describe a simple quantum well, for which interface and barrier contributions to the scattering rate are ignored. Moreover, the screening is only partially accounted for, and  the form factors and overlap integrals that shape the matrix elements of the electron-phonon interaction and screening potential are estimated in the approximation of infinite well barriers. Finally, electron-LO phonon interaction is assumed to be the only mechanism of carrier energy loss, while the crucial role of radiative recombination have been advanced~\cite{Hirst}.
Despite these limitations, the scattering theoretical model we developed for carrier thermalization in 2D systems can provide means of interpretation of why quantum well solar cell has been observed to give rise to a slower cooling rate for carriers than a bulk solar cell in several experimental works reported in the literature. We discuss in this section comparison between our theoretical predictions and available experimental data, which is summarized in terms of quantities that measure hot carrier effect enhancement in quantum well heterostructure in Table 1.

One of the first works comparing quantum well and bulk solar cells did observe a significant reduction of the cooling rate in quantum well solar cells by more than one order of magnitude, by a factor of 60~\cite{Ryan}, which is comparable to our results presented Fig.~\ref{fig:2DThermPower}, where a reduction factor of about 10 is overall obtained.
Many later works suggested that such a reduction of the cooling rate occurs only at a high enough charge density~\cite{Rosenwaks}. In these earlier works, the relative role of confinement, screening, and non-equilibrium hot LO phonon effects have not been addressed in detail.

A pioneering work toward a detailed investigation of the different factors that drive the hot carrier effect in 2D systems was given by Shah and co-workers in Ref.~\cite{ShahPRL}.
They found that the energy loss rate is about 25 times smaller for electrons than for holes  in semiconductor quantum wells (segment on the ratio of thermalization powers for (heavy) hole and electron in Table 1). 
This observation can easily be explained within the framework of kinetic scattering theory in terms of the effective mass dependence of the electron-LO phonon scattering rate. As can be seen from Eq.~(19) of Supp. Mat., the scattering rate depends on the $3/2$ power of the carrier effective mass $m_c$ in the leading order. For GaAs, the heavy hole effective mass $m_h=0.5m_0$ is about 7 times larger than the electron effective mass $m_e=0.07m_0$, where $m_0$ is the electron mass, implying that the cooling rate of holes must be about $7\sqrt{7}\simeq 19$ times larger than that of electrons, providing a reasonable explanation for the observation reported in Ref.~\cite{ShahPRL}. 
Furthermore, the measured energy-loss rate was found to be approximately 8 times smaller than expected for 2D electrons based on existing theoretical calculations at that time~\cite{Ridley2}. This reduction factor is off only by $20\%$ relative to the value obtained from our theoretical results of Fig.~\ref{fig:2DThermPower} b), in which a reduction factor of about 10 is obtained at the minimum of the thermalization power density, where both non-equilibrium LO phonon and screening effects are included (segment on reduction factor in Table 1). In Ref.~\cite{ShahPRL}, this reduction was argued to be caused by non-equilibrium LO phonons whose population exceeds the one expected from equilibrium phonons at lattice temperature, rather than by confinement~\cite{ShahPRL}.

In contrast, our findings gathered in Fig.~\ref{fig:2DThermPower} show that confinement, non-equilibrium LO phonons, and screening effects, all contribute to reducing the electron-phonon scattering rate in quantum well relative to bulk solar cells. 
However, confinement has been concluded to be the principal factor responsible for reducing the scattering rate for the smallest quantum well thicknesses. 
Reducing the system dimensionality exhibits a stronger impact than the one of non-equilibrium LO phonon or electrostatic screening. 
Within the 2D semi-classical framework that we have built, the effect of an a-thermal LO phonon distribution is suppressed due to the confinement-induced reduction of the electron-LO phonon scattering rate which annihilates the acoustic phonon contribution in the LO phonon dynamics, as implied by Eq.~\ref{eq:Neqphonon}. 
Remarkably, this relative feeble role of the non-equilibrium hot LO phonons effect in the carrier thermalization process seems to be in agreement with the result of a classic experiment on GaAs quantum well~\cite{Lyon}.

The use of a multi-quantum well or super-lattice system to enhance efficiency over a bulk GaAs solar cell was first proposed in Ref. \cite{BarnhamDuggan}. The suppression of the radiative recombination in the quantum well compared to that in the bulk regions of the cell was first demonstrated in Ref. \cite{Nelson}. Ref. \cite{EkinsDaukes} demonstrated the excellent material quality of a strain-balanced GaAsP/InGaAs multi-quantum well.  Such material quality enabled the observation of hot carrier effects in these wells in Ref. \cite{Hirst0}. Ref. \cite{Mazzer} proposed that the suppressed radiative recombination in the quantum wells, observed in Ref. \cite{Nelson}, was due to the thermoelectric (Seebeck) effect resulting from the presence of hot carriers in the wells.  

The hot carrier effect has in fact been confirmed in the strain-balanced quantum wells, especially in the so-called ``deep well" samples, for which it was shown that the hot carrier effect is stronger in deep than shallow well samples~\cite{Hirst0}. 
The impact of the quantum well depth can be explained following the arguments pointed out in section~\ref{sec:findings}. Indeed, the deeper the well is, the larger the inter-subband separation energy is, which supports the interpretation that a fewer number of subbands channels thermalization  in the case of  ``deep well" samples, due to the fact that only a fewer number of subbands are able to drive the thermalization process, although a deep well accommodates more subbands.
This interpretation is also in agreement with recent experimental studies on strain-balanced multiple InGaAs/GaAsP quantum wells solar cells, in which it was shown that a deeper quantum well gives rise to significantly higher effective carrier temperature, attributed to hot carrier dynamics~\cite{Hirst}. Indeed, a higher carrier temperature means a weaker system's propensity to loss its energy harvested from light, and thus a lower thermalization power.
Overall, considering deeper wells augments the degree of confinement inside the solar cell absorbers, which is concluded to enhance the hot carrier effect. 

A quantitative measure of this enhancement of the hot carrier effect with a deeper well is given in Table 1 in the segment on well depth dependence. There, we compare a quantity $\eta_{PV}$ which measures how much thermalization power changes relative to the change in the potential barrier height (or equivalently, well depth); $\eta_{PV}<0(>0)$ means lower(higher) thermalization power with deeper well or, equivalently, stronger (weaker) hot carrier effect respectively. The experimental value is estimated from thermalization powers at $T_c=450$K extracted from Fig. 2(a) of \cite{Hirst0} for samples with 20\% In content ($V^{(2)}_c=0.25$eV) and 11\% In content ($V^{(1)}_c=0.17$eV) respectively. The theoretical counterpart values are from our calculation. The thermalization power is found to decrease with deeper well, giving $\eta_{PV}<0$ for all the four theoretical hypotheses, in agreement with the experiment. The experimental reduction factor is about -0.98, closest to our theoretical result with non-equilibrium phonon no screening hypothesis.

\begin{table*}[t]
	\begin{tabular}{l||l|c|l|l}
		Hot carrier effect assessment parameter & Hypothesis  & Our 2D Theory & Experiment \\
		& (2D Theory)     &   (InGaAs/InGaAsP) &    \\
		\hline\hline
		Reduction factor & Eq. NS & 11 & 8(GaAs/AlGaAs)\cite{ShahPRL} \\
		of electron thermalization power $\frac{P^{3D}_{th}}{P^{2D}_{th}}$  & Eq. S  & 20 &  \\
		(evaluated at $L_z=11$ nm as example) & NonEq. NS   & 5   &   \\
		&NonEq. S & 9 & 
		\\
		\hline
		Ratio of (heavy) hole thermalization power  
		& Eq. NS  & 35 & 25(GaAs/AlGaAs)\cite{ShahPRL}\\
		to electron thermalization power  & Eq. S  & 18 &   \\
		Leading-order value: 19 (2D theory) & NonEq. NS   &  35  &   \\
		& NonEq. S & 24 &
		\\
		\hline
		Well depth-dependent enhancement factor  & Eq. NS & -0.66 & -0.98\\
		$\eta_{PV}=\left(\frac{\left(P^{c-LO}_{\mathrm{th}}\right)_2}{\left(P^{c-LO}_{\mathrm{th}}\right)_1}-1\right)/\left(\frac{V^{(2)}_c}{V^{(1)}_c}-1\right)$ & Eq. S  & -4.03  &(GaAsP/InGaAs)\cite{Hirst0}\\
		(evaluated at $L_z=11$ nm as example)  & NonEq. NS  &  -0.85  &     \\
		($V^{(1)}_c=0.235$eV$<V^{(2)}_c=0.4$eV for theory) &NonEq. S & -3.69 & 
		\\
		\hline
		Optimal well thickness $L^*_z$ (nm) & Eq. NS & - & 8.0 (GaAs/AlAs)\cite{ConibeerIEEE} \\
		& Eq. S  &  11.0  &     \\
		& NonEq. NS & - &  \\
		& NonEq. S & 11.5 &
	\end{tabular}
	\caption{Comparison of quantities characterizing hot carrier effect in our theory and available experimental values from slightly different semiconductor materials but still of III-V family. Eq.(NonEq.) and S (NS) respectively represent hypothesis with equilibrium (non-equilibrium) LO phonons and with (without) screening effect. The first segment shows the amount of hot carrier effect enhancement induced by carrier confinement, relative to bulk system. The second segment gives an approximate value of the ratio between thermalization powers of (heavy) hole and electron, first from the leading-order dependence of thermalization power on carriers (electron and (heavy) hole) masses in the first column while giving the exact numerical values at $L_z=11$nm in the third column for various hypotheses. In the third segment, we have defined $\eta_{PV}$ as a measure of the change in thermalization power relative to the change in well depth; $\eta_{PV}<0(>0)$ means a deeper well gives a stronger (weaker) hot carrier effect. In the last segment, an optimal well thickness is defined as the thickness that yields maximum effective carrier (electron) temperature or, equivalently, minimum thermalization power. It is to be noted that this table is not intended to reproduce each available experimental data but to illustrate an overall qualitative and order-of-magnitude quantitative agreement between our theory and experiments.}
	\label{tbl:parameters}
\end{table*}

\begin{figure}
	\centering
	\includegraphics[width=8cm]{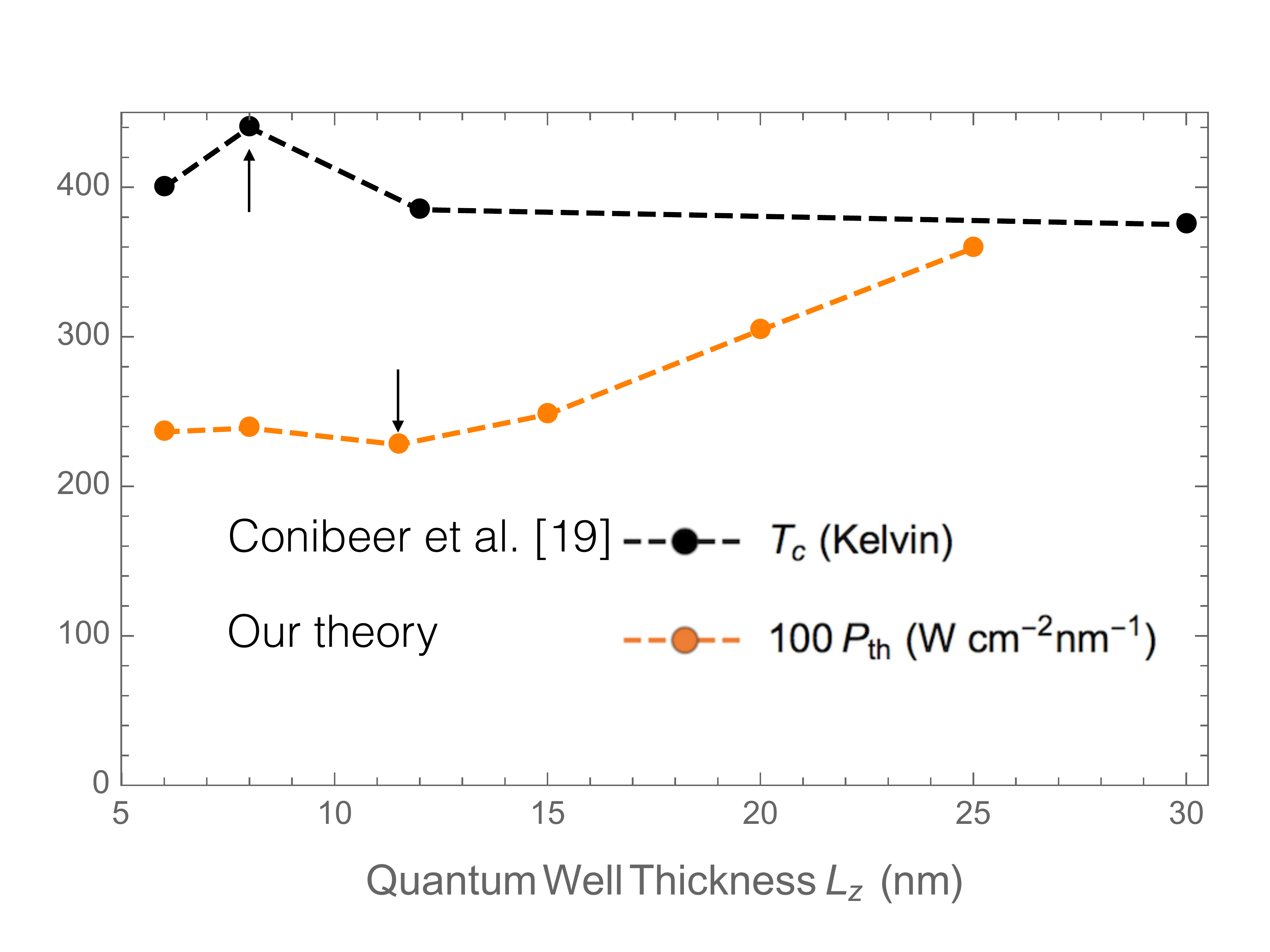}
	\caption{Comparison between experimental curve for carrier effective temperature (black circles) reproduced with permission from Fig. 5 in \cite{ConibeerIEEE} with linear interpolation and our theoretical curve for carrier thermalization power with non-equilibrium phonon and screening hypothesis (orange circles). The arrow in each curve marks the optimum well layer thickness for which the hot carrier effect is maximized. The difference in arrow positions is partly due to different materials used; GaAs/AlAs\cite{ConibeerIEEE} and InGaAs/InGaAsP(ours).}\label{fig:subbandpower}
\end{figure}

Finally, in the investigation conducted by Conibeer et al.~\cite{ConibeerIEEE,ConibeerJJAP}, a non-monotonic dependence of the carrier temperature on the thickness of the quantum well layer was reported, as illustrated in Fig. 9. Phonon emission was considered there as one of the possible mechanisms in explaining the observed hot carrier effect.
While our calculations are based on the parameters of InGaAsP-InGaAs-InGaAsP quantum well layers, we can provide a direct qualitative analysis. 
We argue that the non-monotonic dependence of the carrier temperature on quantum well layer thickness in Ref.~\cite{ConibeerIEEE} corresponds directly to the non-monotonic dependence of the thermalization power density plotted Fig.~\ref{fig:2DThermPower} d) (red solid circles in the inset), within the most realistic framework in which both non-equilibrium LO phonon and screening effects are taken into account, shown in Fig. 9. 
Based on the principle that a lower thermalization power density means a stronger hot carrier effect, the hot carrier effect first increases from 6.5 nm to 10.5 nm, but then decreases from 10.5 nm to 25 nm (and beyond, eventually). 
It suggests that the carrier temperature should increase when going from 6.5 nm to 10.5 nm, and then decreases in going from 10.5 nm to 25 nm (and beyond). This is in accordance with the thickness dependence of the carrier temperature displayed in Fig.~5 of Ref.~\cite{ConibeerIEEE} in going from 6 nm to 30 nm for the quantum well layer thickness, while the barrier thickness is fixed to 40 nm. An optimal thickness of about 11.5 nm in our result and 8 nm in \cite{ConibeerIEEE} gives the strongest hot carrier effect, in terms of lowest thermalization power and highest effective carrier temperature respectively, as summarized in the last segment in Table 1.
Some of the sources of incomplete analysis of this work of Conibeer and co-workers include the facts that we use different materials and that our simple model assumes infinitely thick barriers on both sides compared to the 40 nm thick barrier. 
Overall, bulk 3D theory gives a constant power density, which means that there is no dependence of carrier temperature on the quantum well thickness at all, and thus cannot explain the investigation of Ref.~\cite{ConibeerIEEE}. 

\section{Conclusion.\label{Conc}}

In this work, we have derived a kinetic scattering model for determining the thermalization power in 2D systems, which was used to understand the confinement-induced thermalization reduction in  InGaAsP/InGaAs/InGaAsP quantum wells. In the proposed theoretical investigation, electron and holes are assumed to be in two fixed thermal distributions with fixed temperatures and Fermi level splitting. Moreover, we consider that electron-LO phonon scattering is the unique channel for carrier cooling.
Our main findings are listed below.

1. The thermalization power in quantum wells is reduced by carrier confinement relatively to bulk theory prediction, which occurs because confinement  decreases electron-phonon scattering rate.
Reducing dimensionality is  thus demonstrated to enhance the hot carrier effect, and it is finally expected to improve the efficiency of quantum well solar cells relatively to solar cells with bulk absorbers. 

2. Intra-subband contribution to electron-phonon scattering rate initially dominates over inter-subband contribution at relatively small thicknesses where only few confined electron states exist. As one increases thickness,  inter-subband contribution overtakes that of intra-subband scattering process, as inter-subband terms outnumber intra-subband ones.

3. The effects of confinement, a non-equilibrium LO phonon distribution and screening altogether combine to reduce the electron-phonon scattering and energy loss rates in quantum well heterostructure.
In the 3D case, the scattering rate reduction due to non-equilibrium phonons is more significant than in the 2D case, where the impact of the LO phonon disintegration in two acoustic phonons is indirectly weakened due to the confinement-induced reduction of thermalization. 

4. In the samples studied, it exists a cross-over between the reduction of the first intra-subband contribution and the emergence of both intra and inter-subband contributions of the second subband. It results in a plateau region, where the thermalization power adopts a 3D-like behavior paradoxically originating from electronic state quantization. Including the screening effect then gives rise to a non-monotonic dependence on the well thickness. There is a finite thickness that minimizes the thermalization power density and thus maximizes hot carrier effect. For the InGaAsP-InGaAs-InGaAsP quantum wells under consideration, the optimal thickness is about 10.5 nm. We argue that this optimal thickness corresponds to the optimal thickness that would drive to a maximum value of carrier temperature, as observed experimentally~\cite{ConibeerIEEE}.

Finally, in experiments, there are various other factors that may come into play and contribute to the scattering process or carrier thermalization, such as impurity, defect, interface scattering, thermionic emission, and exciton effects. We leave these complexities as open problems for future works. 
It is to be noted that the various additional complexities not considered in our work tend to provide additional channels to carrier thermalization and thus tend to increase the resulting thermalization power. The numerical value of thermalization estimated from our simple model should thus be interpreted as the lower limit of what one should expect by considering more realistic situations. Furthermore, our semi-classical approach does not discuss interface-type LO phonon modes, which requires more rigorous treatment, such as the one based on dielectric continuum model~\cite{AndoMori}. It is expected that the inclusion of interface-type LO phonon modes would not change qualitatively the conclusions of the present work~\cite{ZhangDysonRidley}. Additionally, surface modes manifest experimentally as a constant additive term on the thermalization power density per surface unit~\cite{GiteauJAP}. Surface modes are thus not expected to change qualitatively the dependence of thermalization power on the thickness of the quantum well (or volume of a bulk sample). 

To conclude, this work provides a proof of principle that 2D confinement of charge carriers in a semiconductor quantum well does indeed give rise to hot carrier effect and precise the details of how this is accomplished, a finding that is expected to be critical and timely to the effort in achieving high solar cell efficiency. 

\textit{Acknowledgements.\textemdash} I.M. acknowledges funding from the French Agence Nationale de la Recherche (ANR) through project ANR-ICEMAN under Grant Number:19-CE05-0019-01. The research is involved within the framework of program 6: PROOF (ANR-IEED-002-02) at IPVF.

	\newpage
	\newpage
	
	\begin{widetext}
		
		\centering
		\textbf{Supplementary Materials: Enhancement of Hot Carrier Effect and Signatures of Confinement in Terms of Thermalization Power in Quantum Well Solar Cell}\\
		\centering
		
		I. Makhfudz$^1$, N. Cavassilas$^1$, M. Giteau,$^2$, $^3$H. Esmaielpour,$^3$, $^4$D. Suchet,$^3$, $^4$A.-M. Dar\'{e} ,$^1$ and F. Michelini $^1$
		
		$^1$ Aix Marseille Universit\'{e}, CNRS, Université de Toulon, IM2NP UMR 7334, 13397, Marseille, France
		
		$^2$Research  Center  for  Advanced  Science  and  Technology, The  University  of  Tokyo,  Komaba  4-6-1,  Meguro-ku,  Tokyo  153-8904,  Japan
		
		$^3$NextPV,  LIA  RCAST-CNRS,  The  University  of  Tokyo, Komaba  4-6-1,  Meguro-ku,  Tokyo  153-8904,  Japan
		
		$^4$CNRS,  Ecole  Polytechnique-IP  Paris,  Institut  Photovoltaique  d'Ile  de  France  (IPVF), UMR  9006,  18  Boulevard  Thomas  Gobert,  91120  Palaiseau,  France

		\date{\today}
		\bigskip
		
		In these Supplementary Materials, we first give the derivation of the generic form of electron-phonon scattering rate. Then, we give the resulting analytical expression for electron-phonon scattering rate when Dirac-delta function approximation is used in place of the exact overlap. The determination of chemical potential starting from quasi-Fermi level splitting and charge neutrality condition in intrinsic semiconductors is then described. 
		Finally, we provide the material parameters used in the calculations.
		
	\end{widetext}
	
\section{Derivation of the Generic Expression for Electron-Phonon Scattering Rate}

The generic expression for the electron-phonon scattering rate reads 
$ \frac{d N_{{\bf q}_\perp,q _z}}{dt} \Bigr|_{c-LO} = \frac{2 \pi}{\hbar} |M_{{\bf q}_\perp,q _z} |^2 A_{\mathbf{q}_{\perp},q_z}$ where
\begin{widetext}
	\[
	A_{\mathbf{q}_{\perp},q_z}=2\sum_{\mathbf{k}_{\perp}}\sum_{k_z,k'_z}I^2_{\mathbf{k},\mathbf{k}-\mathbf{q}}((N_{\mathbf{q}_{\perp},q_z}+1)f_{\mathbf{k}_{\perp},k_z}(1-f_{\mathbf{k}_{\perp}-\mathbf{q}_{\perp},k'_z})
	-N_{\mathbf{q}_{\perp},q_z}f_{\mathbf{k}_{\perp}-\mathbf{q}_{\perp},k'_z}(1-f_{\mathbf{k}_{\perp},k_z}))
	\]
	\begin{equation}\label{TheSum1}
		\times    \delta(E_{\mathbf{k}_{\perp},k_z}-E_{\mathbf{k}_{\perp}-\mathbf{q}_{\perp},k'_z}-\hbar\omega_{\mathbf{q}_{\perp},q_z}) \ .
	\end{equation}
\end{widetext}
We adopt the notation ${\bf k} = ({\bf k}_\perp, k_z)$ and the same for ${\bf q}$. Henceforth we suppose that there is no dispersion in the LO phonon spectrum, thus $\omega_{\mathbf{q}_{\perp},q_z} =\omega_{LO}$.
The difference of carrier energies reads
\begin{equation}
	E_{\mathbf{k}_{\perp},k_z}-E_{\mathbf{k}_{\perp}-\mathbf{q}_{\perp},k'_z}    =\frac{\hbar^2}{2m_c}\left[(k^2_z-{k'_z}^2)-q^2_{\perp}+2k_{\perp}q_{\perp}\cos\phi
	\right] \ ,
\end{equation}
where $\phi$ is the angle between ${\bf k_\perp}$ and ${\bf q}_\perp$.  Using 
\begin{equation}\label{transformationsumtointegral}
	\sum_{\mathbf{k}_{\perp}} ...\rightarrow \frac{S}{(2\pi)^2}\int_0^{2 \pi} d \phi \int_0^{+\infty} d{k}_{\perp} k_\perp \ ... \ ,
\end{equation}
(the upper bound can be safely rejected to infinity), and expressing the discrete nature of the confined electron longitudinal wave vector $k_z=n\pi/L_z,k'_z=n'\pi/L_z$, while leaving the LO phonon longitudinal wave vector $q_z$ as it is, we get
\begin{widetext}
	\[
	A_{\mathbf{q}_{\perp},q_z}=\frac{2 m_c}{\hbar^2 q_\perp} \sum_{n,n'}\frac{S}{(2\pi)^2}
	\int_0^{2 \pi} d \phi \int_0^{+\infty} d{k}_{\perp}
	I^2_{\mathbf{k}_{\perp},n,\mathbf{k}_{\perp}-\mathbf{q}_{\perp},n'}
	\left[(N_{\mathbf{q}_{\perp},q_z}+1)f_{\mathbf{k}_{\perp},n}(1-f_{\mathbf{k}_{\perp}-\mathbf{q}_{\perp},n'})
	-N_{\mathbf{q}_{\perp},q_z}f_{\mathbf{k}_{\perp}-\mathbf{q}_{\perp},n'}(1-f_{\mathbf{k}_{\perp},n})\right]
	\]
	\begin{equation}\label{rateofchangeofbosonnumber1}
		\times \delta\left( \cos \phi +\frac{(n^2-{n'}^2)}{2 k_\perp q_\perp}\frac{\pi^2}{L_z^2} - \frac{q_\perp}{2 k_\perp}  
		- \frac{m_c \omega_{LO} }{\hbar k_\perp q_\perp}
		\right)
	\end{equation}
\end{widetext}
The overlap integral for unconfined phonons $I^2_{\mathbf{k}_{\perp},n,\mathbf{k}_{\perp}-\mathbf{q}_{\perp},n'} = | G_{n,n'}(q_z)|^2 $ was detailed in the main text (Eq.~(21)).
Note that due to the $\delta$-function in Eq.~(\ref{TheSum1}), the arguments of the Fermi functions do not depend on $\phi$ and integration
over angle can be readily made. 
Indeed 
\begin{equation}\label{angularintegral} 
	\int^{2\pi}_0 d\phi\delta(\cos\phi-\cos\phi_0)  =\frac{2}{\sqrt{1-\cos^2\phi_0}} \ , 
\end{equation}
with 
\begin{equation}
	\cos\phi_0= \frac{q_{\perp}}{2k_{\perp}}+\frac{m_c \omega_{LO}}{\hbar k_{\perp}q_{\perp}} - \frac{(n^2-{n'}^2)}{2 k_\perp q_\perp}\frac{\pi^2}{L_z^2} \ .
\end{equation}
This enforces the following constraint, as $k_{\perp}$ and $q_{\perp}$ are
radial wave vectors and so positive, 
\begin{equation}\label{kperpmin} 
	k_{\perp}\geq k_{\perp}^{\mathrm{min}}(n,n') = \Bigl| \frac{q_{\perp}}{2}+\frac{m_c\omega_{LO}}{\hbar q_{\perp}}
	- \frac{(n^2-{n'}^2)}{2 q_\perp}\frac{\pi^2}{L_z^2} 
	\Bigr|  \ ,
\end{equation}
meaning that only electrons from the $n$ subband whose transverse wave vector is greater than $k^{\mathrm{min}}_{\perp}(n,n')$ can emit LO phonons before joining the $n'$ subband.
We have 
\begin{equation}
	\cos\phi_0= \pm \frac{k^{\mathrm{min}}_{\perp}(n,n')}{k_{\perp}}
\end{equation}
giving 
\begin{equation}
	\int^{2\pi}_0 d\phi\delta(\cos\phi-\cos\phi_0)=\frac{2k_{\perp}}{\sqrt{k^2_{\perp}-(k^{\mathrm{min}}_{\perp}(n,n'))^2}}
\end{equation}

Let us turn to the square bracket in Eq.~(\ref{rateofchangeofbosonnumber1}).
The relation imposed by energy conservation ($\delta$-function argument in Eq.~(\ref{TheSum1}))  enables to use the nice identity \cite{Wurfel},
\begin{equation}
	f_{\mathbf{k}_{\perp},n}(1-f_{\mathbf{k}_{\perp}-\mathbf{q}_{\perp},n'}) =N_{\mathbf{q}}(T_c)(f_{\mathbf{k}_{\perp}-\mathbf{q}_{\perp},n'}-f_{\mathbf{k}_{\perp},n}) \ ,
\end{equation}
where $N_{\mathbf{q}}(T_c)=1/(\exp(\hbar \omega_{LO}/(k_BT_c))-1)$ is the equilibrium phonon distribution at the carrier temperature $T_c$.
Gathering the previous results, we finally get
\begin{widetext}
	\begin{equation}\label{aqqz}
		A_{\mathbf{q}_{\perp},q_z}
		=( N_{\mathbf{q}}(T_c) - N_{\mathbf{q}_{\perp},q_z}) \frac{m_c}{\hbar^2 q_\perp} \frac{S}{\pi^2}  \sum_{n,n'} |G_{n,n'}(q_z)|^2
		\int^{+\infty}_{k^{\mathrm{min}}_{\perp}(n,n')} dk_{\perp} \frac{k_{\perp}}{\sqrt{k^2_{\perp}-(k^{\mathrm{min}}_{\perp}(n,n'))^2 }}
		\left[f(E_{\mathbf{k}_{\perp},n}-\hbar \omega_{LO})-f(E_{\mathbf{k}_{\perp},n})\right] \, .
	\end{equation}
\end{widetext}

Finally, defining the scattering time $\tau^{c-LO}_{\bf{q}_\perp,q_z}$ by
\begin{equation}
	\frac{2 \pi}{\hbar} |M_{{\bf q}_\perp,q _z} |^2 A_{\mathbf{q}_{\perp},q_z} = \frac{(N_{\bf{q}}(T_c) -N_{\bf{q}_\perp,q_z})}{\tau^{c-LO}_{\bf{q}_\perp,q_z}}
\end{equation}
and using Eq.~(16) of the main text, we obtain the Eq.~(23) of the body text.

To evaluate the integral on the right hand side in the Eq.~(\ref{aqqz}), namely
\begin{equation}
	I =  \int^{\infty}_{k^{\mathrm{min}}_{\perp}} dk_{\perp}\frac{k_{\perp}}{\sqrt{k^2_{\perp}-(k^{\mathrm{min}}_{\perp})^2}}
	(f_1-f_2) \ ,
\end{equation}
where $f_i = \frac{1}{e^{x_i}+1}$, with respectively $x_1=x_2-\beta \hbar \omega_{LO}$ and $x_2 = \beta (\frac{\hbar^2 }{2 m_c}(k_\perp^2+\frac{n^2 \pi^2}{L_z^2}) -\mu_c)$, 
with $\mu_c$ the carrier chemical potential and 
$\beta = 1/k_B T_c$, we can use  
an obvious change of variable which  leads to
\begin{equation}
	I = \frac 1 2 \frac{\sqrt{2 m_c k_B T_c}}{\hbar}  \int_0^{+\infty} dx \frac{1}{\sqrt{x}} \Bigl[ \frac{1}{e^{(x-a)}+1} -\frac{1}{e^{(x-b)}+1}  \Bigr] \ ,
\end{equation}
with
\begin{eqnarray}
	a&=& b +\beta \hbar \omega_{LO}\, , \rm{and} \\
	b &=& -\beta \Bigl[\frac{\hbar^2 }{2 m_c}((k^{min}_\perp(n,n'))^2+\frac{n^2 \pi^2}{L_z^2}) -\mu_c\Big]\, .
\end{eqnarray}
Eqaution (14) can be expressed in terms of a polylogarithmic function using Ref.~\onlinecite{nist},
\begin{equation}
	\int_0^{+\infty} \frac{dt}{\sqrt{t}} \frac{1}{e^{t-u}+1}= - \Gamma \Bigl(\frac 1 2 \Bigr) Li_{1/2} (- e^u) \ .
\end{equation}
This leads to
\begin{equation}
	I =  \frac 1 2 \frac{\sqrt{2 m_c k_B T_c}}{\hbar} \Gamma \Bigl( \frac 1 2\Bigr ) (Li_{1/2}(- e^{b}) -Li_{1/2}(-e^{a}) ) \, ,
\end{equation}
and finally, the expression for the electron-phonon scattering rate takes the form
\begin{eqnarray}\label{eqSM:taumc}
	\frac{1}{\tau^{c-LO}_{\bf{q}_\perp,q_z}}
	&=&\frac{ 2S}{h}   \frac{|M_{{\bf q}_\perp,q _z} |^2 }{ q_\perp}
	\Big[ \frac{2m_c}{\hbar^2} \Big]^{3/2}  \sqrt{ k_B T_c}    \sum_{n,n'} |G_{n,n'}(q_z)|^2 \nonumber \\ 
	&& \times \Gamma \Bigl( \frac 1 2\Bigr ) (Li_{1/2}(- e^{b}) -Li_{1/2}(-e^{a}) ) \, .
\end{eqnarray}

\section{Scattering Rate for Confined Phonons} 

There is a substantial literature concerning the electron-LO-phonon interaction in quantum wells, particularly about the confrontation between macroscopic dielectric models with various boundary conditions and microscopic models \cite{Rudin91,Rucker91}. The former are easier to manipulate,
while the latter maybe more quantitative. In Ref.~\cite{Rucker91} it was shown that crude bulk models may provide acceptable results.  
In the main part of the paper we have considered unconfined phonons, by arguing that barriers and quantum well lattice parameters are quite close. In 
the present section we consider the opposite case of confined phonons \cite{Ridley89}, however ignoring any interface phonon. 
In this case, the overlap (Eq.~(21) of the main text) will be replaced by
\begin{widetext}
	\begin{equation}
		G_{n, n'}(m) = \int^{L_z}_0 dz \psi_{n}(z)\cos(q_z z)\psi^*_{n'}(z) =\frac 1 2 \Bigl( - \delta_{m, n+n'} 
		+  \delta_{m, n'-n} +\delta_{m,n-n'}
		\Bigr) \ ,
	\end{equation}
\end{widetext} 
where the longitudinal phonon wave number satisfies $q_z = \frac{ m \pi}{L_z}$. 
Aside from the values of $|G_{n, n'}(q_z)|^2$, the calculations are the same, and we can reuse present Eq.~(\ref{aqqz}) and Eq.~(23) of the main text.

\begin{figure*}[ht!]
	\includegraphics[width=8cm]{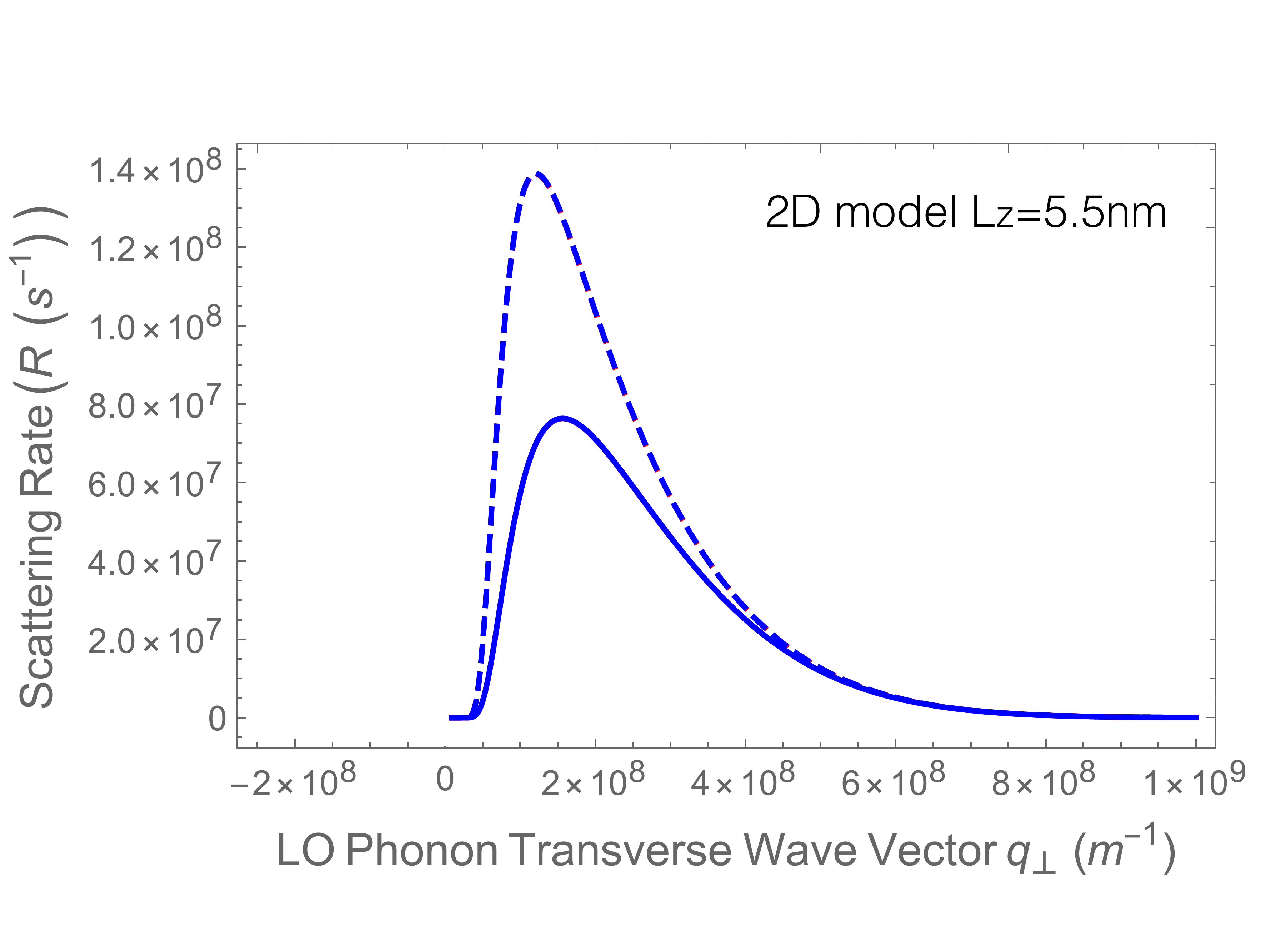}
	\includegraphics[width=8cm]{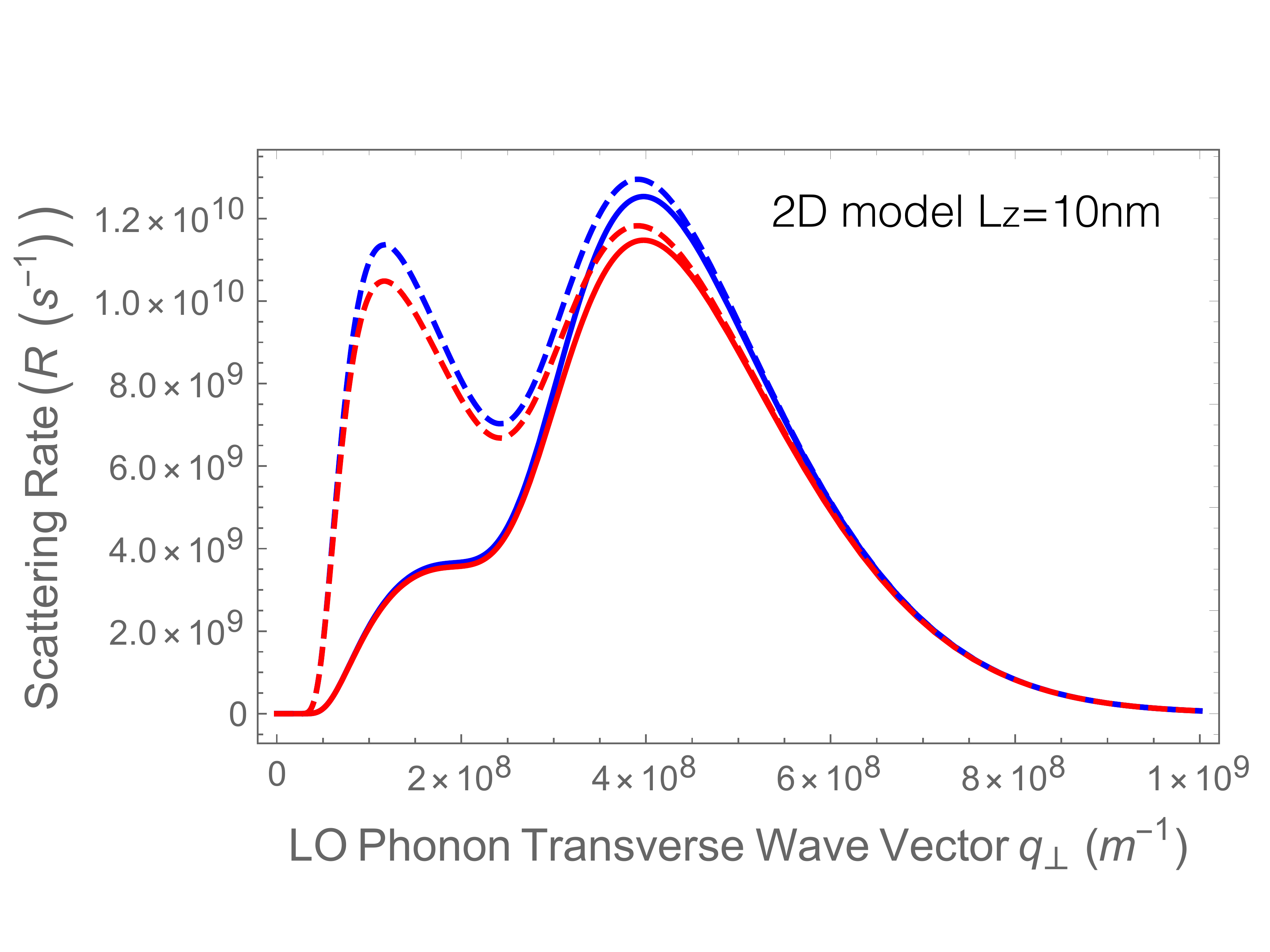}
	\includegraphics[width=8cm]{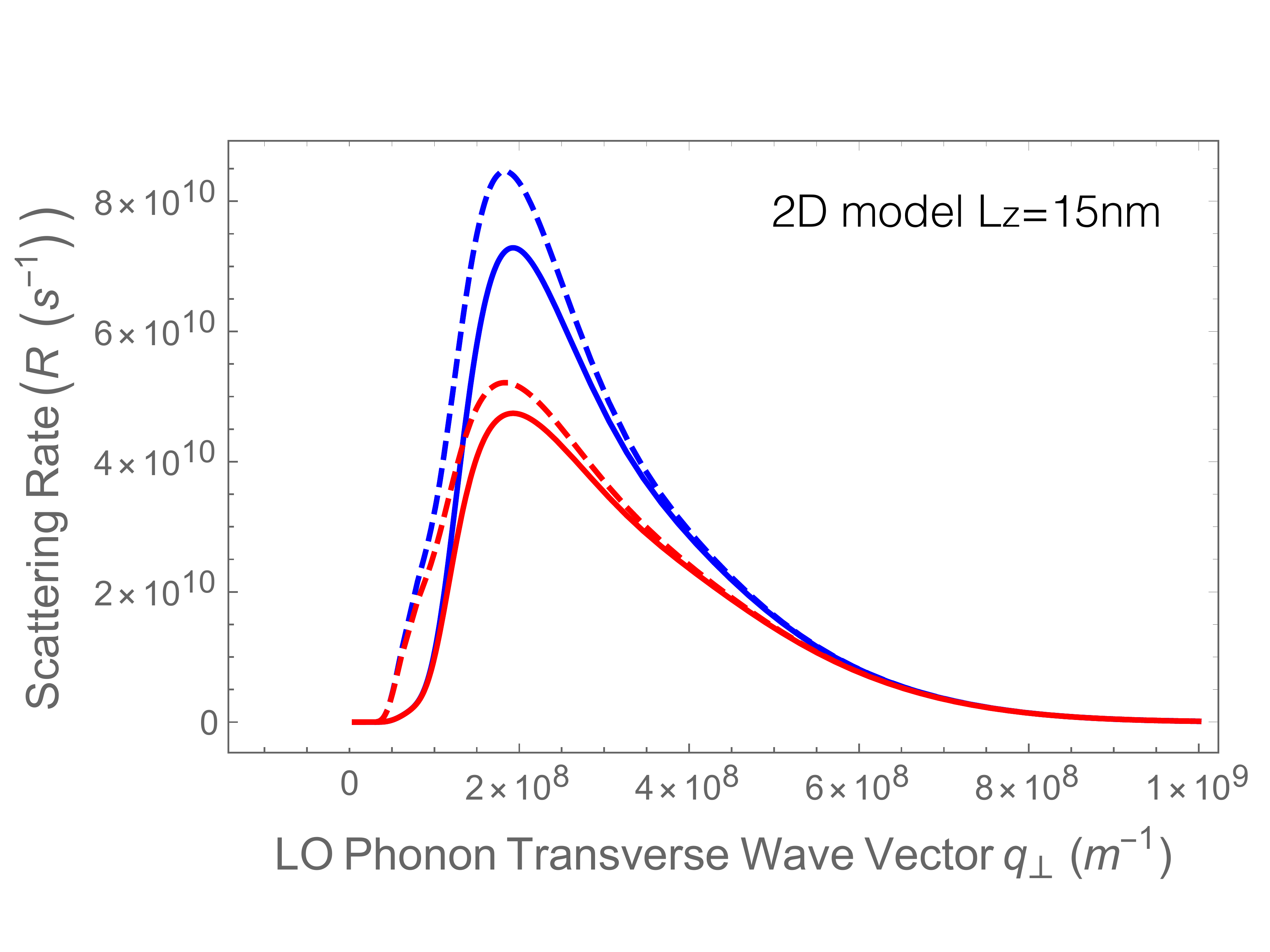}
	\includegraphics[width=8cm]{3Dscreenedvsnotscreened.pdf}
	\caption{ \label{ConfinedPhonons} 2D Electron-LO confined phonon scattering rates with screening (solid lines) and without screening (dashed curves), summed over $q_z$, as functions 
		of transverse wave vector $q_{\perp}$ for different well widths, as well as replicated 3D results (lower right quadrant) from Fig. 5 of the main text.}
\end{figure*}

We plot in Fig.~\ref{ConfinedPhonons}, for the quantum well, the quantities
\begin{equation}
	\sum_m \frac{1}{\tau^{c-LO}_{\bf{q}_\perp, m}+(\tau^{LO-ac}_{\bf q})}
\end{equation}
$\tau^{LO-ac}_{\bf q}$ in parenthesis being present or absent respectively in the non-equilibrium and equilibrium cases. With the same convention, the 3D results out-of and at equilibrium, 
$1/(\tau_{{\bf q}}^{c-LO} +(\tau^{LO-ac}_{\bf q}))$ are also presented.
Results with and without screening are also shown.
Contrarily to the unconfined phonon case, (Fig.~5 of the body text), in quasi 2D, for all three thicknesses, the resulting confined LO phonon scattering rate is lower than $1/\tau^{LO-ac}_{\mathbf{q}}\simeq 1.36\times 10^{11}\mathrm{s}^{-1}$, in such a way that the effect of non-equilibrium state of LO phonons does not make any visible difference for $L_z = 5.5$ nm, and is still faint for $L_z=10$~nm. 
In this case, LO phonons are close to equilibrium with their acoustic counterparts at lattice temperature.
Finally, we notice that for $L_z=10$~nm, phonon confinement causes two peaks to appear on the curve of the scattering rate. These peaks are actually related to the phonon modes involved by the thermalization selection rules. For $L_z=5.5$~nm, there is only one band and therefore only one peak, while for $L_z=15$~nm, the overlapping of the peaks does not allow to separate them.

\begin{widetext}
	\begin{figure*}[ht!]
		\includegraphics[width=9cm]{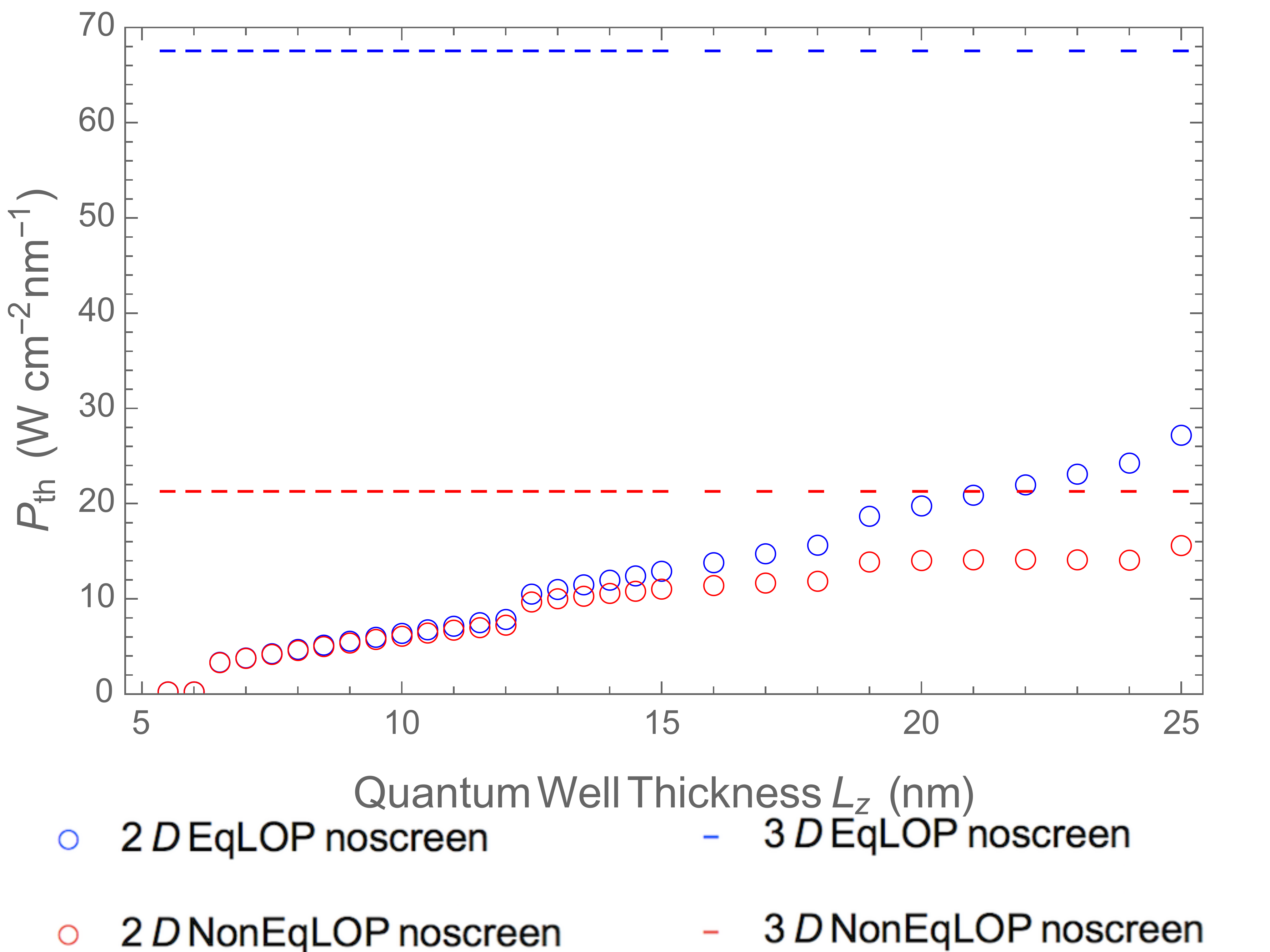}\includegraphics[width=9cm]{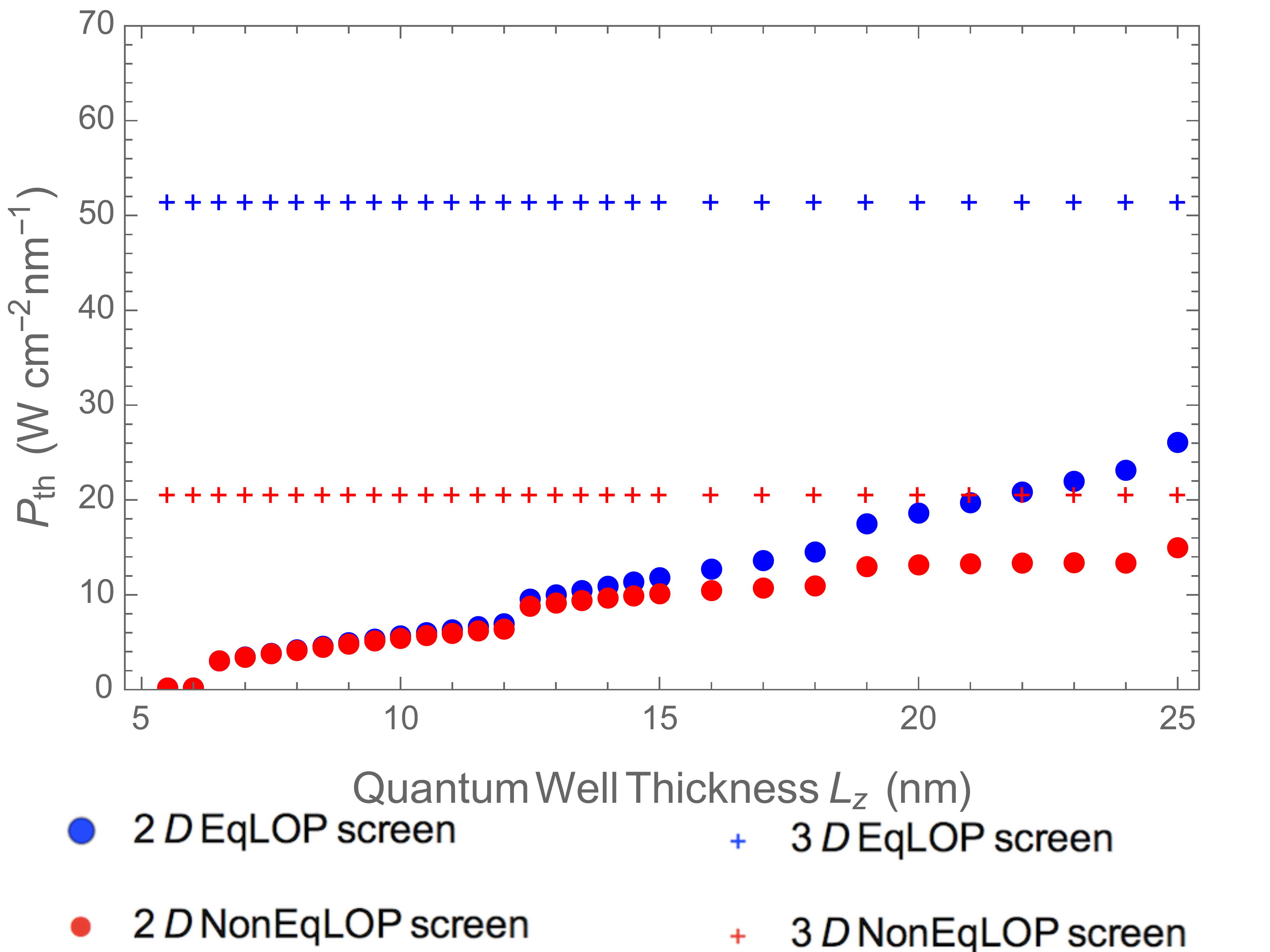}
		\includegraphics[width=9cm]{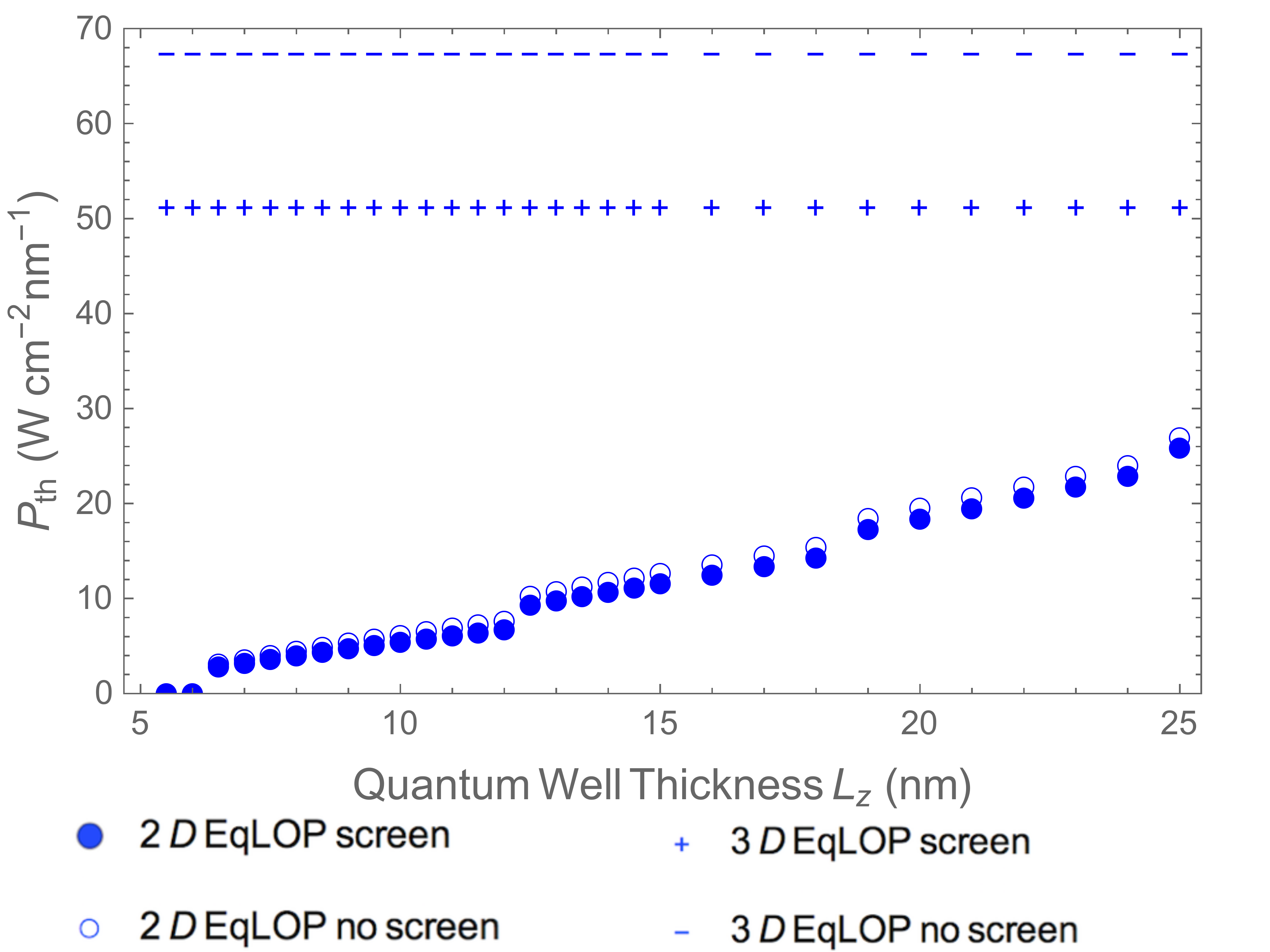}\includegraphics[width=9cm]{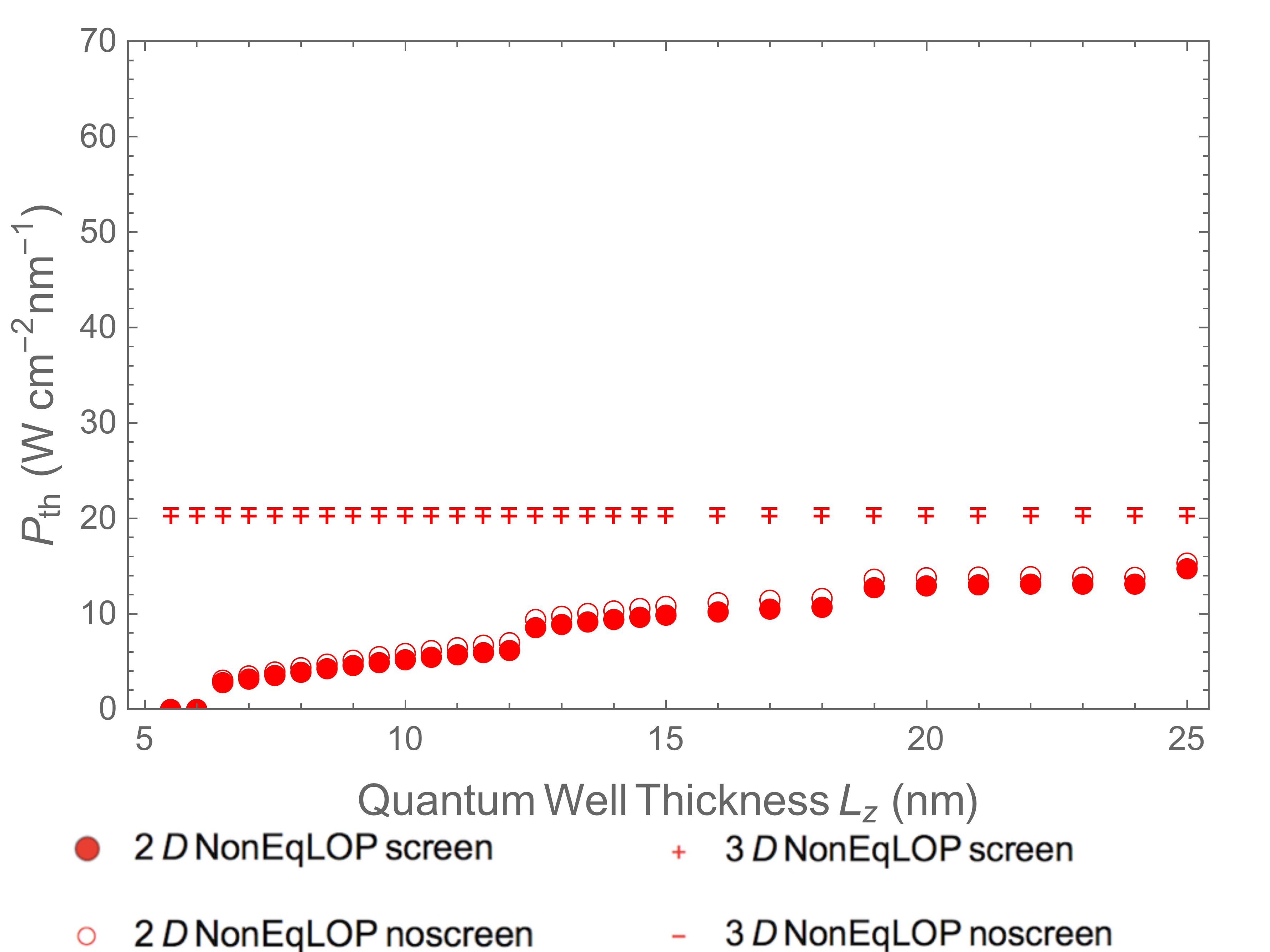}
		\caption{\label{ThermPow} Thermalization power per unit volume as a function of quantum well thickness, considering all possible combinations of assumptions: with and without non-equilibrium LO phonons (upper two figures, while the screening is included or not) and with and without screening effect (lower two figures, while the LO phonons are assumed in non-equilibrium or equilibrium states), computed with $\Delta \mu=0.653$~eV, $T_c=450$~K, and $T_L=300$~K. }
	\end{figure*}
\end{widetext}

The resulting thermalization power for confined phonons is presented in Fig.~\ref{ThermPow}, where the results for various combination of the state of LO phonons (equilibrium or non-equilibrium) and screening (included or not) are displayed.
The curves show that 2D thermalization power always falls below the 3D results for the same state of LO phonon and screening.
The reduction in thermalization power is rather strong for the first two quantum well layer thicknesses (5.5 and 6 nm) where there is only one confined electron state, by about two orders of magnitude. However, once we increase the quantum well layer thickness to accomodate two or more confined electron states, the degree of reduction becomes only about one order of magnitude, closer to those found in the existing works in the literature, e.g. \cite{MonteCarloS}. The 2D curves also exhibit well defined jumps in the thermalization 
power density as one increases the thickness, reflecting the quantization of energy levels of confined electrons. 

\section{Determination of the chemical potential}

The quantity which connects our theory to photoluminescence experiment, is the chemical potential $\mu_c$. Indeed the quasi-Fermi level splitting, that is the difference between electron and hole chemical potentials, can be measured experimentally. 
To determine chemical potentials, we assume that electrons become hot at temperature $T_c$, and so do the light holes, while the heavy holes remain cold at lattice temperature $T_L$, in tune with the finding of transient study~\cite{JoshiFerry}, due to a cooling rate which is slower for light holes than for heavy holes. 
We have verified that considering the same temperature for heavy and light holes does not change our results qualitatively. We substitute these effective temperatures $T_L$ and $T_c$ into charge neutrality equation when solving for the carrier chemical potential as described in detail below.    

The chemical potential $\mu_c$ used in the calculations is determined from solving the charge neutrality equation for intrinsic semiconductors, taking into account light and heavy holes which share the same chemical potential but not the same temperature:
\begin{equation}\label{chargeneutralityintrinsic}
	n_c=p_{vl}+p_{vh} \ ,
\end{equation}
where $n_c$ is the electron density, and $p_{vl} (p_{vh})$ are respectively light (heavy) hole densities.
In 3D, the electron density is defined as
\begin{equation}
	n_c = \int_{\varepsilon_c}^{+\infty} d \epsilon g_c (\epsilon) \frac{1}{e^{(\epsilon-\mu_c)/k_B T_c}+1} \ ,
\end{equation}
where $\varepsilon_c$ is the conduction bande edge. 
Using well known results for 3D density of states, we have 
\begin{equation}
	n_c = \frac {1}{2 \pi^2} \Bigl( \frac{2 m_c k_B T_c}{\hbar^2} \Bigr)^{3/2} \int_0^{+\infty} dx \frac{\sqrt{x}}{e^{x-\eta_c}+1} \ ,
\end{equation}
with $\eta_c = (\mu_c -\varepsilon_c)/k_B T_c$. While, for holes, from the definition
\begin{equation} 
	p_v = \int_{-\infty}^{\varepsilon_v} d\epsilon g_h(\epsilon) \frac{1}{e^{(\mu_h-\epsilon)/k_B T_h}+1} \ ,
\end{equation}
with $\varepsilon_v$ the valence band edge, we find
\begin{equation}
	p_{v} = \frac {1}{2 \pi^2} \Bigl( \frac{2 m_{v} k_B T_{h}}{\hbar^2} \Bigr)^{3/2} \int_0^{+\infty} dx \frac{\sqrt{x}}{e^{x+\eta_h}+1} \ .
\end{equation}
$m_v$, $T_h$ and $\eta_h$ need to be specified for both hole types:
For light holes (mass $m_{hl}$, temperature $T_c$), we have $\eta_{hl} = \eta_c+(E_g-\Delta \mu)/k_B T_c$ and the corresponding density will be noted $p_{vl}$, while, for heavy holes (mass $m_{hh}$, temperature $T_L$), 
$\eta_{hh}= \frac{T_c}{T_L}\eta_c + (E_g-\Delta \mu)/k_B T_L$, and the density will be noted $p_{vh}$.
We have defined $E_g=\varepsilon_c -\varepsilon_v$, the band gap, and $\Delta \mu =\mu_c -\mu_h$ the quasi Fermi level splitting. 
The implicit Eq.~(\ref{chargeneutralityintrinsic}) enables to determine $\eta_c$, hence $\mu_c -\varepsilon_c$.

In quasi 2D, for finite depth quantum well 
of width $L_z$, the electron density of state is expressed as a sum over the subbands given by
\begin{equation}
	g_{c,Q2D} (\epsilon)= \frac{m_c}{\pi \hbar^2}\frac{1}{L_z}  \sum_{i=1}^\infty \Theta(\epsilon-(E_{ic}+\varepsilon_c)) \ ,
\end{equation}
with $E_{ic}$ the edges of the different subbands.  It leads to
\begin{equation}
	n_c = \frac{m_c k_B T_c}{\pi \hbar^2 L_z} \sum_i \int_0^{+\infty} d x \frac{1}{e^{x+ E_{ic}/k_BT_c-\eta_c}+1} \ ,
\end{equation}
where $\eta_c$ has been previously defined. The integral can be evaluated, this gives
\begin{equation}
	n_c = \frac{m_c k_B T_c}{\pi \hbar^2 L_z} \sum_i \ln{(1+e^{\eta_c -E_{ic}/k_BT_c})} \ .
\end{equation}
For holes, we use the following DOS
\begin{equation}
	g_{v,Q2D} (\epsilon)=\frac{m_h}{\pi \hbar^2} \frac{1}{L_z} \sum_i \Theta \Bigl( \varepsilon_v -E_{ih}-\epsilon \Bigr)
\end{equation}
with $E_{ih}$ the edges of the different hole subbands.
For light holes (mass $m_{hl}$, temperature $T_c$), the calculation gives
\begin{equation}
	p_{vl} =   \frac{m_{hl}}{\pi \hbar^2} \frac{k_B T_c}{L_z} \sum_i \ln{\Bigl(1+e^{-\eta_{hl}- E_{ihl} /k_BT_c}\Bigr)} \ .
\end{equation}
On the other hand, for the heavy holes (mass $m_{hh}$, temperature $T_L$), we get
\begin{equation}
	p_{vh} =   \frac{m_{hh}}{\pi \hbar^2} \frac{k_B T_L}{L_z} \sum_i \ln{\Bigl(1+e^{-\eta_{hh}- E_{ihh} /k_BT_L}\Bigr)} \ .
\end{equation}
$\eta_{hh}$ and $\eta_{hl}$ have been formerly defined.

The quasi-Fermi level splitting $\Delta\mu$ and the carrier temperature are determined from the modeling of photoluminescence spectrum, which measures the photon current density and can be described by the generalized Kirchoff law~\cite{Wurfel}:
\begin{equation}\label{photoncurrent}
	j_{\gamma}(\hbar\omega)=a(\hbar\omega)\frac{(\hbar\omega)^2}{4\pi^2\hbar^3c^2}\frac{1}{\exp(\frac{\hbar\omega-\Delta\mu}{k_BT_c})-1}
\end{equation}
where the absorptivity is given by
\begin{equation}
	a(\hbar\omega)=\frac{\left[1-R(\hbar\omega)\right](1-\exp(-\alpha(\hbar\omega)L_z))}{1-R(\hbar\omega)\exp(-\alpha(\hbar\omega)L_z)}\, ,
\end{equation}
$R$ being the reflectivity and $\alpha$ the absorption coefficient. 
In the proposed investigation, we assume the quasi-Fermi level splitting $\Delta\mu$ and the electron temperature to be fixed as we vary the quantum well thickness, even in the limit of the 3D case. Translated into experimental conditions, it would mean that we have discussed the thermalization power in different quantum well heterostructures, tuning the excitation conditions to provide the same photoluminescence spectrum.

\begin{figure}
	\includegraphics[width=\columnwidth]{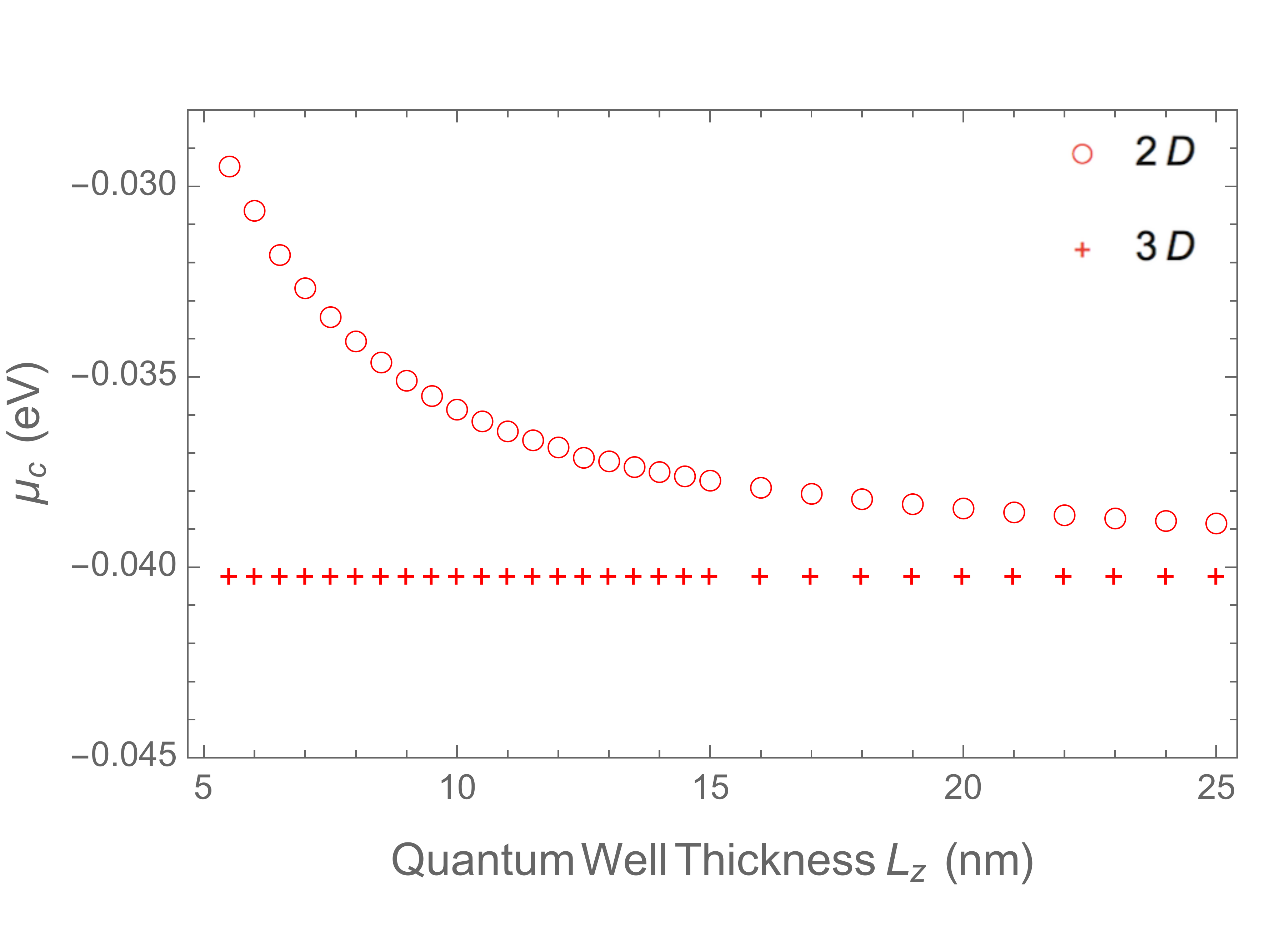}
	\label{figpotchim}
	\caption{Profiles of the chemical potential for our 2D model (circle) and 3D model (+), for quantum well layer thicknesses $L_z=5.5-25$ nm.}
\end{figure}

In the calculations, we have used the data from experiment using 405 nm laser irradiating multiple quantum well absorber consisting of 5 InGaAsP-InGaAs-InGaAsP layers with thickness 10nm-5.5nm-30nm and transverse dimension $L_{\perp}=3$~cm: $\Delta \mu=0.653$~eV and $T_c=450$~K.
From these data, we have estimated the chemical potentials from the 3D model and the proposed 2D model, together shown in Fig.~\ref{figpotchim}, where the asymptotic convergence towards the 3D case is nearly reached in the quasi-2D case beyond 15~nm.

\section{Material Parameters}

Material parameters that were used to produce the figures of the main text and the Suppl. Mat. are gathered in Tab.~\ref{tbl:parameters}.

\begin{table}
	\begin{tabular}{l|c|l}
		Quantity & Symbol & Value \\
		\hline
		LO phonon energy & $\hbar\omega_{LO}$ & $36$ meV \\
		Energy gap (InGaAs) & $E_g$ & 0.8 eV\\
		Well depth for electrons & $V_{con}$ & 0.235 eV\\
		Well depth for holes & $V_{val}$ & 0.185 eV\\
		Electron effective mass (InGaAs) & $m_c$ & 0.041 $m_0$ \\
		Heavy hole effective mass & $m_{hh}$ & 0.363 $m_0$\\
		Light hole effective mass & $m_{hl}$ & 0.051 $m_0$\\
		Sound velocity (InGaAs) & $v_s$ & $3.5\times 10^3$ ms$^{-1}$ \\
		Infinite-frequency susceptibility & $K_{\infty}$ & 12.9\\
		Static susceptibility & $K_s$ & 10.9\\
		Deformation potential coefficient &  $\Gamma$ & 0.43\\
		LO phonon decay time (InGaAsP) &  $\tau^{LO-ac}_0$ & $2.2\times 10^{-11}$ s\\
		(0 K)& \\
		Lattice constant (InGaAsP) &  $a$ & 5.75 \AA\\
		Material density (InGaAsP) &  $\rho$ & 5500 kg/m$^3$ \\
		\hline
	\end{tabular}
	\caption{\label{tbl:parameters} Material parameters needed to compute the electron-LO phonon scattering rate and thermalization power in InGaAsP/InGaAs/InGaAsP quantum well heterostructures of varying thickness.}
\end{table} 


\end{document}